\documentclass[aps,prd,nofootinbib,showpacs]{revtex4}

\usepackage{amsmath,amsfonts,amssymb}
\usepackage{graphicx}

\def\eq#1{{Eq.~(\ref{#1})}}

\begin{document}

\title{Non-relativistic limit of quantum field theory in inertial and non-inertial frames and the Principle of Equivalence}
\author{Hamsa Padmanabhan}
\email{hamsa.padmanabhan@gmail.com}
\affiliation
{Department of Physics, University of Pune, Ganeshkhind,
Pune 411 007, India}
\author{T. Padmanabhan}
\email{paddy@iucaa.ernet.in}
\affiliation{IUCAA, Pune University Campus, Ganeshkhind,
 Pune 411 007, India}

\begin{abstract}
We discuss the non-relativistic limit of quantum field theory in  an inertial frame, in the Rindler frame and in the presence of a weak gravitational field, and attempt to highlight and clarify several subtleties. In particular, we study the following issues: (a) While the action for a \textit{relativistic} free particle is invariant under the Lorentz transformation, the corresponding action for a \textit{non-relativistic} free particle is not invariant under the Galilean transformation, but picks up extra contributions at the end points. This leads to an extra phase in the non-relativistic wave function under a Galilean transformation, which can be related to the rest energy of the particle \textit{even in the non-relativistic limit}. We show that this is closely related to the peculiar fact that the relativistic action for a free particle remains invariant even if we restrict ourselves to $\mathcal{O}(1/c^2)$ in implementing the Lorentz transformation.  (b) We provide a brief critique of the principle of equivalence in the quantum mechanical context. In particular, we show how solutions to the generally covariant Klein-Gordon equation in a noninertial frame, which has a time-dependent acceleration, reduce to the non-relativistic wave function in the presence of an appropriate (time-dependent) gravitational field in the $c\to\infty$ limit,
and relate this fact to the validity of the principle of equivalence in a quantum mechanical context. We also show that the extra phase acquired by the \textit{non-relativistic} wave function in an accelerated frame, actually arises from the gravitational time dilation and survives in the non-relativistic limit. (c) While the solution of the Schrodinger equation can be given an interpretation as being the probability amplitude for a single particle, such an interpretation fails in quantum field theory. We show how, in spite of this, one can explicitly evaluate the path integral using the (non-quadratic) action  for a relativistic particle (involving a square root) and obtain the Feynman propagator. Further, we describe how this propagator reduces to the standard path integral kernel in the non-relativistic limit. (d) We show that the limiting procedures for the propagators mentioned above work correctly even in the presence of a weak gravitational field, or in the Rindler frame, and discuss the implications for the principle of equivalence. 
\end{abstract}

\maketitle

\section{Introduction}

\begin{figure}
 \begin{center}
  \includegraphics[scale=0.4]{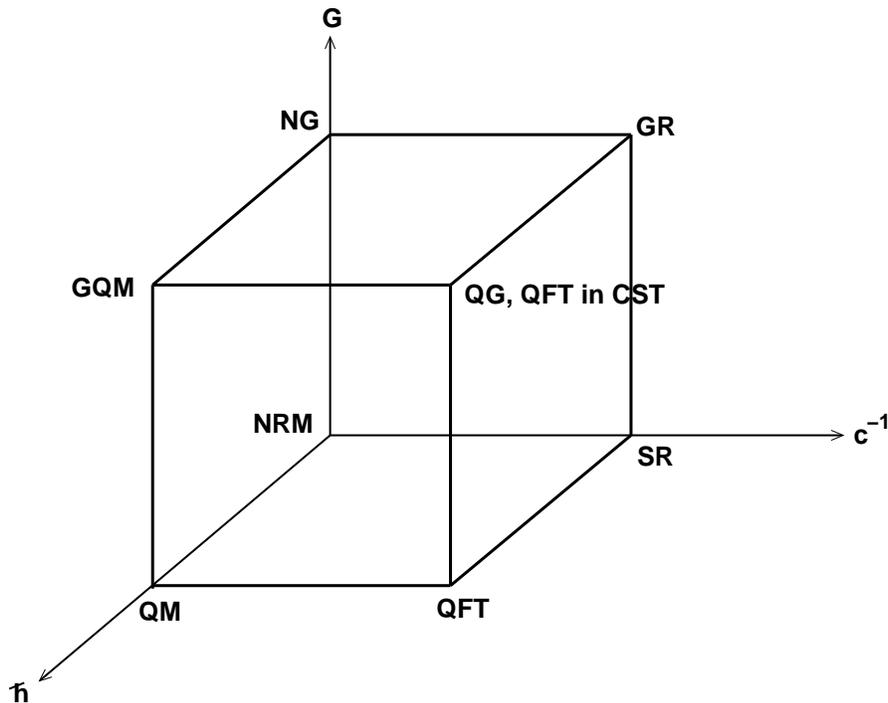}
 \end{center}
\caption{The ``physics cube''. See text for discussion.}
\label{fig:physicscube}
\end{figure}

The fundamental principles of physics, as we understand them today, emphasize the role of three constants: $G$ (Newton's gravitational constant), $c$ (the speed of light) and $\hbar$ (the Planck constant). By a suitable choice of units, we can set the numerical value of all these three to unity and the broad structure of physical theories can be represented using   a 3-dimensional space in which each of the Cartesian coordinates is taken to be one of the above mentioned fundamental constants (see Fig. \ref{fig:physicscube}). It turns out to be convenient to use $(1/c)$ rather than $c$ in such a description, and the entire space of physical theories will be confined within the unit cube so formed.

The examination of this diagram reveals several interesting features. The origin  $G=0, \ \hbar=0, \ c=\infty$ represents an idealized non-relativistic (point) mechanics (NRM) with which many physics courses begin. Moving along the speed of light axis to $c^{-1} =1$, (keeping $G=0, \hbar =0$), will get us special relativistic (SR) mechanics. Similarly, moving along the $G-$axis will lead to non-relativistic, classical, Newtonian gravity (NG) and travelling along the $\hbar-$axis will lead to non-relativistic quantum mechanics (QM). More exact theoretical structures emerge when a \textit{pair} of constants are non-zero. The vertex $c^{-1}=1, \ G=1$, \ $\hbar=0$ represents classical general relativity (GR) which combines the principles of special relativity and gravity. Similarly, $\hbar =1, \ c^{-1}=1$, $G=0$ leads to  flat spacetime quantum field theory (QFT) which combines the principles of special relativity and quantum theory.
The vertex at which  all the three constants are unity, $c^{-1}=1,\ G=1, \ \hbar=1$, should represent the domain of quantum gravity but, more importantly for our purpose, \textit{it also represents the study of quantum field theory in curved space-time} (QFT in CST), like, for example, the study of radiation from black holes. (A description of the thermal features of  black holes requires all these three constants to be non-zero.) While quantum gravity still remains a distant dream, we do have a fair amount of understanding of quantum field theory in curved spacetime and, in this sense, this vertex (QFT in CST) can be considered as within our grasp.

While most of  these limiting forms of physical theories have attracted a reasonable amount of attention and made it into textbooks, the above  diagram brings out one limiting case which probably  has not been explored in comparable detail. This is the vertex with $c^{-1}=0, \ G=1, \ \hbar=1$, which corresponds to exploring the  nature of gravity in a quantum \textit{mechanical} context (GQM). Much of the discussion in this paper is devoted to the exploration of this vertex
and, more generally, to projecting the theories to the $G\hbar$  plane by taking the $c\to \infty$ limit in different contexts.\footnote{One of the authors (TP) has used this diagram during his lectures in the mid-eighties. A somewhat similar diagram with a tetrahedron rather than a cube appears in \cite{kuchar}. It is very likely that many others have thought of such a description but we could not find  a published reference.
}

 This vertex is of interest for several reasons.  
 First, there are interesting questions related, for example, to the principle of equivalence and the free-fall of atomic systems in Newtonian gravitational fields which are  conceptually and experimentally important \cite{urev,gqm}. This topic has received a fair amount of attention in the context of neutron interferometry \cite{neutron} as well as in the context of neutrino physics \cite{dharam}. There are some controversies (and possible lack of clarity) in the literature on this topic, which makes it  worth studying. 
 
 Second, this vertex serves as an interesting test-bed for several limiting processes. For example,  note that one can arrive at the GQM vertex from the quantum field theory in curved spacetime (QFT in CST) vertex  by taking the limit $c\to \infty$. It will be interesting to ask what limiting forms the known results in quantum field theory in curved spacetime take when we  project (move) along this direction and do a Taylor series expansion in inverse powers of $c$.  (For example, it would be interesting to ask what is the lowest non-trivial order in $c^{-1}$ one should work at to reproduce, say, the thermal nature of the inertial vacuum.) 
 
 Third, this limit also has interesting implications even in the context of $G=0$, viz., when we move from quantum field theory to quantum mechanics. How does  quantum \textit{field} theory reduce to \textit{single particle} quantum mechanics when we take the $c\to \infty$ limit, and how does this limiting process get modified when we ``switch on'' $G$? We will see that there are  some subtleties even in the simplest context of transition from special relativity to non-relativistic mechanics because the symmetry groups --- the Lorentz group and the Galilean group --- have very different structures \cite{galilean}. 
 
 Finally, it is interesting to investigate the role of the principle of equivalence in quantum mechanical and field theoretic phenomena. (This issue has attracted attention in many of the papers quoted in \cite{urev,gqm}). We know that one can mimic the local effects of gravity by a  coordinate transformation to 
a suitable non-inertial frame. Therefore, a corresponding transformation to a non-inertial frame in quantum mechanics must allow us to incorporate the effects of gravity and take us from the QM to the GQM vertex. It is interesting to verify this explicitly and understand the details.

The key new results in our paper are the following:

(i) We show that the reason for the non-invariance of the Galilean action under Galilean transformations is intimately connected to the existence of \textit{rest-mass energy of the particle which leaves a residue even in the non-relativistic limit}. The details of the calculation might help to clarify some controversial conclusions in the literature
\cite{greenberger}. 

(ii) We point out that the special relativistic action is invariant under a Lorentz transformation \textit{even if we ignore terms of order $1/c^4$ and beyond}, and discuss the consequences of this result, which does not seem to have been noticed previously in the literature.

(iii) We show that the principle of equivalence does hold good in quantum mechanics, \textit{if} it is interpreted as embodying the equivalence of the \textit{dynamical equations} in an accelerated frame and in a gravitational field. We explicitly prove this equivalence for the case of the Schrodinger equation in an inertial and an accelerated frame of reference.

(iv) We show that the relativistic Feynman propagator in quantum field theory can be obtained from a path integral over $\exp(i A/\hbar)$, just as in the case of non-relativistic quantum mechanics, but now   with the standard special relativistic action $A$  \textit{which is not quadratic}. We also show explicitly \textit{how} the  Feynman propagator reduces to the Feynman path-integral kernel in the non-relativistic limit, not only in the inertial frame, but also in the Rindler frame, and in the more general context of a weak gravitational field. This procedure also allows us to obtain the leading relativistic \textit{correction} to the Feynman path-integral. 

(v) We show that the reason for the \textit{non-transitivity} property of the Feynman propagator is the same as that for the non-transitivity of the corresponding \textit{energy kernel} in the non-relativistic theory, and we elaborate and discuss this connection. 

Throughout the paper, we have tried to be as self-contained as possible by giving the necessary pedagogical details.

\section{Behaviour of systems under Lorentz transformations and Galilean transformations}

We begin by discussing the --- apparently elementary --- situation of the transition from special relativity  to non-relativistic mechanics (NRM) by taking the limit $c^{-1}\to 0$ which involves moving along  SR to NRM in Fig. \ref{fig:physicscube}. While  text books consider this limiting procedure as straightforward, we will see that some curious features arise when we consider the limiting form of the action functional in this context. (Not surprisingly, this issue has led to some controversial conclusions in published literature \cite{greenberger} which we also hope to clarify.) As is well known,  special relativity is invariant under a Lorentz transformation of the coordinates, while non-relativistic mechanics is invariant under a Galilean transformation of the coordinates. Given the fact that one recovers the Galilean coordinate transformation by setting $c=\infty$ in the Lorentz transformation equations, one would have thought that any theory which is Lorentz invariant will lead to a theory which is invariant under Galilean transformations in the limit of $c\to \infty$. As we shall see, there are several subtleties in the manner by which Galilean invariance of the theory arises in this limit from Lorentz invariance. This topic is usually discussed in the literature in relation to the structure of the Galilean group and superselection rules in quantum mechanics (see e.g., \cite{galilean}). We shall explore the issues in a more straightforward and transparent manner.

\subsection{Behaviour of action functionals under Lorentz and Galilean transformations}

We begin by enquiring how the action functional for a free particle changes under Lorentz transformations in special relativity and Galilean transformations in non-relativistic mechanics. We note that the action $\mathcal{A}$  is given by  
\begin{equation}
 \mathcal{A} = \int{L(\mathbf{x},\mathbf{v},t) dt}
\end{equation} 
where the Lagrangian $L$ can be expressed as a function $L = L(v^2)$ of the square of the magnitude of the particle's velocity, by taking into account the homogeneity of space and time and the isotropy of space.
This holds  both in the case of relativistic and non-relativistic mechanics.

It is impossible to proceed further and determine  the explicit form of the Lagrangian, without making some additional assumptions.  We now have to make a distinction between non-relativistic and relativistic mechanics by  postulating  the invariance of physical laws under different sets of coordinate transformations.

In non-relativistic theory, we  postulate that the equations of motion should retain the same form when we make a 
 Galilean transformation:
\begin{equation}
\label{Galilean}
 x = x' + V t; \  t = t'
\end{equation}
from the co-ordinates $(x,t)$ of an inertial frame $S$ to the co-ordinates $(x', t')$ of another frame $S'$ moving with a uniform velocity $V$ along the positive $x-$direction with respect to $S$. (In this and what follows, we suppress the two spatial dimensions and work in (1 + 1)-dimensions for simplicity.) 
The corresponding velocity transformation is $
 v = v' + V$
where  $v$ and $v'$ are the velocities measured in frames $S$ and $S'$ respectively.

The sufficient (though not necessary) condition for 
the equations of motion to retain the same form in both $S$ and $S'$ is that the action should be invariant under the transformations in \eq{Galilean}. It is straight forward to see that \textit{no} non-trivial function $L(v^2)$ has this property and hence \textit{we fail to construct} an action functional which remains invariant under the Galilean transformation. 

The usual procedure at this stage is to note that the equations of motion will remain invariant even if the Lagrangian is not, as long as the Lagrangian changes only by the addition of a total time derivative of a function of coordinates and time.
It is again easy to show that  this requires 
 the Lagrangian for the free particle to be \textit{proportional} to the square of the velocity i.e. $L \propto v^2$ or $L = (1/2) m v^2$ where $m$ is defined to be the mass of the particle. In this case, the Lagrangians $L$ and $L'$ in the frames of reference $S$ and $S'$ respectively are related by:
\begin{equation}
\label{laggalchange}
 L = \dfrac{1}{2}m v^2 = \dfrac{1}{2}m (v' + V)^2  = L' + \dfrac{d}{dt}(mx'V + \dfrac{1}{2}m V^2 t)
\end{equation} 
Hence, in the non-relativistic theory, the Lagrangians $L$ and $L'$ differ by a total time derivative of a function of co-ordinates and time. The corresponding actions differ by contributions at the end points:
\begin{equation}
 \mathcal{A}= \mathcal{A}' + \left.\left(mx'V +\frac{1}{2} mV^2t'\right)\right|_1^2
 \label{diffact}
\end{equation} 
Note that the canonical momentum  ($p' = p - mV$) and the energy $(E'= E-pV+(1/2) m V^2)$ are \textit{not} invariant when we transform from $L$ to $L'$.

In special relativity, we replace \eq{Galilean}
by the Lorentz transformations  between $S$ and $S'$ of the form:
\begin{equation}
 x = \dfrac{x' + V t'}{(1 - V^2/c^2)^{1/2}}; \quad  t = \dfrac{t' + V x'/c^2}{(1 - V^2/c^2)^{1/2}}
 \label{lt}
\end{equation} 
The transformation of velocities is now given by:
\begin{equation}
 v = \dfrac{v' + V}{1 + v' V/c^2}
\end{equation} 
In the limit of $c\rightarrow \infty$, the above equations reduce to the corresponding ones for the Galilean transformation. 

Curiously enough, it is now possible to construct an action functional which \textit{is actually invariant} (instead of picking up an extra boundary term) under the transformations in \eq{lt}. This is given by
\begin{equation}
 \mathcal{A} = \alpha \int (1 - v^2/c^2)^{1/2} dt
 \label{alphact}
\end{equation} 
where $\alpha$ is a constant. The above form of the action remains invariant under Lorentz transformations.  It is \textit{not} possible to choose $\alpha$ such that \eq{alphact} reduces to the action for non-relativistic mechanics when $c\to \infty$. The best we can do is to choose $\alpha$ in such a way that in the non-relativistic limit, we get back the non-relativistic form of the action, \textit{apart from a constant term} in the Lagrangian. This amounts to setting $\alpha = -mc^2$. Hence, in special relativity, the action for a free particle is taken to be:
\begin{equation}
 \mathcal{A} = -mc^2 \int (1 - v^2/c^2)^{1/2} dt
 \label{relaction}
\end{equation} 
The above discussion raises a couple of questions which 
require investigation.

First, we expect the non-relativistic theory to arise as a limiting case of the fully relativistic theory, in the limit of $c\rightarrow \infty$. It is true that the Lorentz transformation equations reduce to the Galilean transformation equations in the limit of $c\rightarrow \infty$. However, the special relativistic action does \textit{not} reduce  to the non-relativistic action in this limit, but instead picks up an extra term, $-mc^2t$ evaluated at the end points. As we shall see, this term --- which is usually ignored in textbooks as being due to ``just an addition of a constant to a Lagrangian'' ---  has some interesting implications for the structure of special relativity and non-relativistic mechanics. This is already hinted at by the fact that the relativistic action in \eq{relaction}
blows up in the limit of $c\to \infty$ and does not have a valid limit in the strict mathematical sense. If one can ``renormalize'' this action by adding a term $\mathcal{A}_1 \equiv mc^2 t$ to \eq{relaction}, then $\mathcal{A} + \mathcal{A}_1$ will have a proper limit. But then, $ \mathcal{A}_1$
is \textit{not} Lorentz invariant, and hence this ``renormalization'' is not a valid procedure. We will see repeatedly that the term $mc^2 t$ plays a crucial role in our discussions.

Second, we find that in the special relativistic theory, which is the more exact theory, the action remains invariant under special relativistic (Lorentz) transformations. However, in the non-relativistic limit, the corresponding non-relativistic action does \textit{not} remain invariant under the analogous non-relativistic (Galilean) transformations, but again picks up some extra terms. This is surprising, when seen in the light of the Galilean transformation being a limiting case of the Lorentz transformation.\footnote{As an aside, we note the following amusing fact: We implemented homogeneity in time and space in the free particle Lagrangian by excluding the explicit dependence of $L$ on $\textbf{x}$ or $t$ but incorporated Galilean invariance by allowing $L$ to pick up a total derivative. It is possible to do the converse. One can write down free particle Lagrangians which are \textit{strictly invariant} under Galilean transformations but depend explicitly on $t$ and $\textbf{x}$ through a total time derivative. A simple example is $L=(1/2)m(\textbf{v}-\textbf{x}/t)^2$ which is \textit{invariant} under a Galilean transformation but depends on $t$ and $\textbf{x}$ through a total time derivative; $L$ differs from $L_0=(1/2)mv^2$ by the total time derivative $-d/dt((1/2)mx^2/t)$.} 

Both the above issues are  usually ignored by noting that the equations of motion do not change when a total time derivative of a function of coordinates and time is added to the Lagrangian, and hence such action functionals are equivalent as far as physical phenomena are concerned. 
This is true in \textit{classical} physics, but in \textit{quantum} theory, the form of the action is closely related to the phase of the wave function. Our analysis suggests that the phase of the wave function of a free particle remains invariant under Lorentz transformations, but in the $c\to \infty$ limit, this invariance gets broken. We will now explore this situation by studying the phase of the free particle wave function in relativistic quantum theory and its limiting form in non-relativistic quantum mechanics. 

\subsection{Behaviour of wave functions under Lorentz and Galilean transformations}
\label{subsec:lorgal}

The key issue  can be introduced by examining the following argument which seems to allow one to find the wave function of a \textit{moving} free particle  from that of a \textit{stationary} free particle. The argument runs as follows: Consider a particle of mass $m$ which is at rest in a frame $S'$ that is moving with a velocity $V$ along the $x-$axis of another inertial frame $S$.
We postulate that (a) the phase of the wave function in $S'$  should depend on the rest energy $ mc^2$ in the usual manner and hence the wave function should be $\psi \propto \exp(-imc^2t')$. If we also assume that (b)  $\psi$ is a scalar under Lorentz transformations, then the wave function in $S$ is given by
\begin{equation}
 \psi\propto \exp(-imc^2t')\propto \exp[-imc^2\gamma (t-Vx/c^2)] \propto \exp[-iEt+ipx]
 \label{relpsi}
\end{equation} 
where the symbols have their usual meanings and $E=\gamma mc^2$ and $p=\gamma mV$ are the relativistic energy and momentum of the particle in the frame $S$ in which it is moving with velocity $V$. Thus, given the postulates (a) and (b) mentioned above, we can determine the wave function of a moving particle from that of a particle at rest using the Lorentz transformation.\footnote{Strictly speaking, quantum field theory does not allow the notion of a single-particle wave function with a probabilistic interpretation. We shall discuss this issue more rigorously in Section \ref{sec:relnonrelprop}, but the key points mentioned here continue to hold even when we take this fact into account. Hence we will not worry about it in this Section.}

If we now take the limit of $c\to \infty$, we find that $p\to mV$ and $E\to mc^2 + p^2/2m$. Hence the wave function of a non-relativistic particle seems to be given by
\begin{equation}
 \psi_{\rm NR} \propto \exp[-imc^2 t] \exp[-i(1/2)mV^2t+imVx]
\end{equation} 
Thus we  seem to have obtained the correct solution to free particle Schrodinger equation ``except for'' the phase term $(-imc^2t)$ which has no interpretation in non-relativistic mechanics. Again, strictly speaking, $\psi$ does not have a limit when $c\to \infty$. One needs to consider, instead, the quantity $\psi \exp[imc^2t]$ to get the correct limit, but, just as in the case of the action functional, this is not a valid prescription. The correct $\psi$ in a relativistic theory is indeed the one in \eq{relpsi}
and not the one
with an $mc^2t$ term subtracted out from the phase.

The situation becomes more curious when we realize that we  \textit{could not have done any of these} using Galilean transformations alone. That is, 
there  is no way of obtaining the wave function of a moving particle from that of a stationary particle by the  application of a Galilean transformation.
To begin with, the postulate (a) above has no meaning in non-relativistic quantum mechanics and so we cannot write down a wave function with a time dependent phase in its rest frame. Second, as we shall see, wave functions do not transform as ``scalars'' under Galilean transformations.  To get everything right, one has to add an extra phase, the origin of which we will now discuss.

Let us again start with the phase of a relativistic wave function $(Et-px)$. We know that under a Lorentz transformation, this goes over to
\begin{equation}
 (Et-px) \Longrightarrow (E' t' - p'x')
\end{equation} 
where $E'\equiv\gamma (E-Vp)$; $p'\equiv\gamma (p-VE/c^2)$. In this transformation, we treat $E,p$ as just two real numbers parameterizing the wave function, and are \textit{not} assuming that they are components of a four-vector, etc.    But once we have applied the Lorentz transformation \textit{to the coordinates,} we can interpret the coefficients of $t',x'$ (which are $E',p'$)
as the energy and momentum of the particle in $S'$, and this interpretation, as we know, is correct.

None of this works with Galilean transformations and non-relativistic quantum mechanics. Under a Galilean transformation, the phase of the free particle wave function in non-relativistic quantum mechanics transforms as 
\begin{equation}
 \frac{p^2}{2m} t - px \Longrightarrow \left(\frac{p^2}{2m} - pV\right)t' - px'
 \label{trans1}
\end{equation} 
which does not allow us to read off the coefficients of $t'$ and $x'$ as the physical energy and momentum of the non-relativistic particle in $S'$. So, unlike in the relativistic case, we cannot identify the correct energy and momentum from the phase of the wave function after a Galilean transformation. We can, of course, introduce by hand the two parameters $p'=p-mV, E'=E-Vp+(1/2)mV^2$ and re-express the right hand side of \eq{trans1} in terms of $p',E'$.  The result is best presented in terms of the identity:  
\begin{equation}
\left( \frac{p'^2}{2m} t'-p'x'\right) -\left( \frac{p^2}{2m}t -px\right) =   \left( mVx - \frac{1}{2} mV^2 t\right) = \left(mVx' + \frac{1}{2} mV^2 t'\right)
\label{neweqn}
\end{equation}
So, if we express the phase in $S'$ in terms of $E'$ and $p'$, then the phase is not invariant and we need to add an extra phase in the right hand side of \eq{neweqn}.
This extra phase is essentially the difference between the two actions in \eq{diffact}.
 Since the phases of wave functions are classical actions for a free particle, it is clear that the non-invariance of the action under Galilean transformations reflects itself in the additional phase that arises in the wave function.

These facts are also  related to the contrasting  behaviour of the Klein-Gordon  equation and the Schrodinger equation under Lorentz transformations and Galilean transformations, respectively.  With future applications in mind, we will consider the transformation of the Schrodinger equation under  a more                                                                                                                                                                                                                                                       general  co-ordinate transformation from a frame of reference $S = (t, x)$ to a frame  $S' = (t, x') \equiv (t, x - \xi(t))$ where $\xi(t)$ is some arbitrary function of time. 
(The Galilean transformation is a special case when $\xi(t) = Vt$; when $\xi(t)$ is not a linear function of $t$ this represents a transformation to a non-inertial frame.)
Using the fact that $(\partial/\partial x)_t = (\partial/\partial x')_{t'}$, $(\partial/\partial t)_x
=(\partial/\partial t')_{x'} - \dot \xi (\partial/\partial x')_{t'}$ we can transform the free particle Schrodinger equation
\begin{equation}
 i\frac{\partial\Psi}{\partial t} = - \frac{1}{2m} \frac{\partial^2\Psi}{\partial x^2} 
\end{equation} 
into the $(t',x')$ coordinates and obtain
\begin{equation}
  i\frac{\partial\Psi}{\partial t'}=- \frac{1}{2m} \frac{\partial^2\Psi}{\partial x'^2} 
  +i\dot \xi \frac{\partial\Psi}{\partial x'}
\end{equation} 
(We have set $\hbar =1$ for convenience.)
In the case of the Galilean transformation, we have $\xi(t) = Vt$,  leading to
\begin{equation}
  i\frac{\partial\Psi}{\partial t'}=- \frac{1}{2m} \frac{\partial^2\Psi}{\partial x'^2} 
  +iV \frac{\partial\Psi}{\partial x'} 
  \label{gtschrod}
\end{equation} 
Obviously the free particle Schrodinger equation is not form invariant under the Galilean transformation if we consider $\Psi$ to be a scalar --- unlike the Klein-Gordon equation, which is form invariant, with the wave function remaining a scalar. It is the second term on the right hand side of \eq{gtschrod} which requires an extra phase to be added to the wave function for everything to be consistent.

Once again, all this is closely related to the fact that the action in special relativity is invariant under Lorentz transformations while the action in non-relativistic mechanics picks up an additional term under Galilean transformations. To see this explicitly, we will again consider the behaviour of the \textit{action} for a free particle under the transformation from a frame of reference $S = (t, x)$ to a frame  $S' = (t, x') \equiv (t, x - \xi(t))$ as before, and relate it to the extra phase in the wave function explicitly. The Lagrangian of the free particle in $S'$ is given by 
\begin{equation}
L' = (1/2) m \dot{x'}^2   = (1/2) m \dot{x}^2 - m \dot{x}\dot{\xi} + (1/2) m \dot{\xi}^2                                                                                                                                                                                                                                                                                                                                                                                                                                     \end{equation} 
which can be written in the form
\begin{equation}
\label{lagtrans}
 L' = L + \dfrac{df}{dt} 
\end{equation} 
where
\begin{equation}
\label{fform}
f \equiv -mx\dot{\xi} + \int (1/2) m \dot{\xi}^2 dt                                                 
\end{equation} 
and $L$ is a new Lagrangian: 
\begin{equation}
 L = (1/2) m \dot{x}^2 + mx\ddot{\xi}
\end{equation}
which is equivalent to $L'$ as far as the equations of motion are concerned, since the total time derivative $df/dt$ does not contribute to the equations of motion. This $L$ represents the Lagrangian for a particle acted upon by a force $m\ddot{\xi}$ or, equivalently, a particle located in a spatially homogeneous but time dependent gravitational field  $\ddot{\xi}$. (In fact, the entire phase $f$ acquires a simple physical meaning when we take gravitational time dilation into account; we will say more about this in Section \ref{sec:qmequivalence}.)

We recall that when  we add $d f(x,t)/dt$, to the Lagrangian $L$, thus transforming it to $L' = L + df/dt$, both the canonical momentum and the Hamiltonian  change, becoming $p' = p + \partial f/\partial x$ and $H' = H - \partial f/\partial t$ respectively. 
In quantum mechanics, the time evolution of the wave function is determined by the Hamiltonian operator and hence,  the form of the wave function must change when we make a co-ordinate transformation from the frame $S$ to the frame $S'$.
Let $\Psi'(t, x')$ be the quantum-mechanical wave function for the free particle in the frame $S'$. Then, it can be shown that  the corresponding wave function $\Psi(t, x)$ for the same particle in the frame $S$ is given by:
\begin{equation}
\label{wavefntrans}
\Psi(t, x) = \Psi'(t, x - \xi(t)) e^{-if/\hbar}
\end{equation} 
and it satisfies the equation
\begin{equation}
\label{schrotrans}
i \hbar  \dfrac{\partial\Psi(t, x)}{\partial t} = -\dfrac{\hbar ^2}{2 m}\dfrac{\partial^2\Psi(t, x)}{\partial x^2} - m\ddot{\xi} x \Psi(t, x)
\end{equation} in the frame of reference $S$, which is just the Schrodinger equation with a ``pseudo-potential'' energy term, $-m\ddot{\xi} x$. This term arises because the frame $S$ is an accelerated frame of reference and  has a close link with the principle of equivalence, which we shall take up later on in Section \ref{sec:qmequivalence}. (This result has been derived numerous times in the literature; for completeness we reproduce the derivation in Appendix \ref{app2}).

Coming back to the case of Galilean transformation, we have $\xi(t) = Vt$ and 
\begin{equation}
 f=-mxV + \frac{1}{2} mV^2 t.
\end{equation} 
Therefore, \eq{wavefntrans} takes the form
\begin{equation}
\label{wavefntrans1}
\Psi(t, x) = \Psi'(t, x - Vt) \exp[(-i/\hbar) (-mxV + (1/2) mV^2 t)]
\end{equation}
That is, we need to transform the wave function, treating it as a scalar, \textit{and then add an extra phase} which is consistent with what we found in \eq{neweqn}.
As we have said before, all this is perfectly consistent as regards the application of the Galilean transformation in quantum mechanics.

How is it that the Klein-Gordon equation is invariant under the Lorentz transformation, but the Schrodinger equation --- which is presumably obtained in the $c\to \infty$ limit of the Klein-Gordon equation ---  is \textit{not} invariant under the Galilean transformation, given the fact that the Lorentz transformation reduces to the Galilean transformation in the appropriate limit?

This has to do with the manner in which one obtains the Schrodinger equation from the Klein-Gordon equation and brings to the center stage the role of the $mc^2$ term in the phase. We will outline how the extra phase in \eq{wavefntrans1} can be obtained from  a fully invariant Klein-Gordon equation. Consider the wave function $\Phi (t,x)$ which is the solution to a free particle Klein-Gordon equation. We know that under a Lorentz transformation, $\Phi (t,x)\Longrightarrow \Phi' (t',x')$,  thus transforming as a scalar. To obtain the Schrodinger equation for a wave function $\psi(t,x)$ we first have to separate the $mc^2 t$ term from the phase of the $\Phi$ by writing 
\begin{equation}
 \Phi(t,x) = \psi(t,x) \exp[-imc^2 t]
\end{equation} 
It is then straightforward to show that in the limit  of $c\to \infty$, $\psi(t,x)$ will satisfy a free particle Schrodinger equation. [We will demonstrate a more general result in the presence of a gravitational field later on, and hence we skip the algebraic details here; see \eq{gravschrodeqn}.] To obtain the Schrodinger equation in $S'$, we have to similarly write $\Phi'(t',x')=\psi'(t',x') \exp(-imc^2t')$. The fact that $\Phi$ transforms as a scalar can now be used to relate $\psi$ and $\psi'$, and we find that 
\begin{equation}
 \psi' = \psi\exp[-imc^2(t-t')]
\end{equation} 
We see that, in addition to the scalar transformation, the wave function picks up a phase which is just $mc^2(t-t')$. 
\textit{Incredibly enough, this expression has a finite, nonzero limit when $c\to \infty$!}
Evaluating this quantity in the limit of $c\to \infty$, we get 
\begin{eqnarray}
 mc^2 (t-t') &=& mc^2 \gamma \left( t' + \frac{Vx'}{c^2}\right) - mc^2 t'\nonumber\\
 &=& mc^2 \left[ t' + \frac{Vx'}{c^2} + \frac{1}{2} \frac{V^2 t'}{c^2} + \mathcal{O}\left( \frac{1}{c^4}\right)\right] - mc^2 t'
\nonumber\\
&=& mVx' + \frac{mV^2 t'}{2} +\mathcal{O}\left( \frac{1}{c^2}\right)
\label{wavephase}
\end{eqnarray} 
This is precisely the mysterious phase which occurs in the Schrodinger equation under a Galilean transformation. It has a simple interpretation as being equal to $mc^2(t-t')$, thus  emphasizing the role of rest energy \textit{even in the non-relativistic limit}. We will see later that this interpretation holds even in the presence of gravity, if we take the gravitational time dilation effect into account.

\subsection{A little surprise: Invariance under truncated Lorentz transformation}\label{sec:invunderlt}

There is another peculiar result closely related to the one obtained above which deserves mention. We have already seen that 
the action for a relativistic particle is invariant under the Lorentz transformation while the action for the non-relativistic particle is not invariant under the Galilean transformation. But it turns out --- somewhat surprisingly ---  that  the action for the relativistic particle remains invariant even if we only retain the terms in the action upto and including $\mathcal{O}(1/c^2)$ (and ignore terms of $\mathcal{O}(1/c^4)$ and higher) and also treat the Lorentz transformation to the same order of approximation.

This may be seen as follows: In the expression for the relativistic action  for a free particle, we expand the integrand in a power series retaining terms upto and including order $1/c^2$, and ignoring all higher order terms. We then find: 
\begin{eqnarray}
 \mathcal{A} &=& -mc^2 \int (1 - v^2/c^2)^{1/2} dt 
 \nonumber \\&=& -mc^2\int \frac{dt'}{\sqrt{1 - (V^2/c^2)}} \sqrt{1 - (V^2/c^2)- (v'^2/c^2) + \mathcal{O}(1/c^4)}
 \nonumber\\ &=& -mc^2\int dt' \left(1 + \frac{V^2}{2 c^2} + \mathcal{O}(1/c^4) \right) \left(1 - \frac{v'^2}{2 c^2} - \frac{V^2}{2 c^2} + \mathcal{O}(1/c^4) \right) 
 \nonumber\\ &=&  -mc^2\int dt' \left(1 - \frac{v'^2}{2 c^2} + \mathcal{O}(1/c^4) \right)
 \label{quadaction}
\end{eqnarray} 
However, to the same order of approximation, we also know that:
\begin{equation}
 \mathcal{A} =  -mc^2\int dt \left(1 - \frac{v^2}{2 c^2} + \mathcal{O}(1/c^4)\right)
 \label{apprrelaction}
\end{equation} 
Hence we find that $\mathcal{A}$ and  $\mathcal{A'}$ are identical in form, even to this order in $(1/c)$! (This is not at all an obvious result and it occurs due to the cancellation of the $(V^2/2c^2)$ term coming from the velocity transformation, with that coming from the $t'$ to $t$ transformation.) Clearly, when both these transformations are carried out consistently, the action remains form-invariant, even if we disregard $\mathcal{O}(1/c^4)$ and higher terms!

Note that the action in \eq{quadaction} is what a text book would have considered as being the ``same as the action in non-relativistic mechanics except for a constant added to the Lagrangian''.  But we have proved that the action for a non-relativistic particle is \textit{not} invariant under a Galilean transformation, but picks up an extra term.
If we express the change in the non-relativistic action under the Galilean transformation as $\Delta \mathcal{A}_G$, we can write:
\begin{equation}
\label{deltaagal}
 \Delta \mathcal{A}_G = m\int dt\left[\frac{1}{2} v^2 - \frac{1}{2} v'^2 \right] =  mc^2 \int dt \left(1 - \frac{v'^2}{2 c^2}\right) - mc^2 \int dt \left(1 - \frac{v^2}{2 c^2}\right)
\end{equation} 
We see that this difference $\Delta \mathcal{A}_G$ bears a close resemblance to the difference between the corresponding relativistic action functionals, with the crucial difference being that the \textit{time has been kept invariant} in \eq{deltaagal} while the expressions in \eq{quadaction} and \eq{apprrelaction} have integrals over $t'$ and $t$. This shows clearly that the reason for the non-invariance of the Galilean transformation lies in equating $t$ to $t'$. If we take into account the difference between $t$ and $t'$, even to the lowest order in $(1/c^2)$, it will lead to invariance of the non-relativistic action functional.

The invariance of the approximate action in \eq{quadaction} allows us to obtain the result
\begin{equation}
 mc^2 (t - t') = \frac{1}{2} m \left[ \int dt\ v^2 -\int dt' v'^2\right]
\end{equation} 
The left hand side can be interpreted as the difference in the phase of a wave function arising due to the rest mass energy in two coordinate systems. We see that this quantity is equal to the expression on the right hand side which is completely independent of $c$! In other words, this phase difference which occurs in a wave function through the factor $\exp[-imc^2 (t-t')]$ survives to a \textit{completely non-relativistic expression} in the $c\to \infty$ limit as already seen in \eq{wavephase}. This result, which is surprising at first sight, arises from the fact that the difference between the coordinate time and the proper time carried by a moving clock is given by 
\begin{equation}
 \Delta \tau = t - \int dt \ (1 - v'^2/c^2)^{1/2} \approx \frac{1}{2c^2} \int v'^2 dt
\end{equation} 
where the last equality holds to $\mathcal{O}(1/c^4)$. It follows that the phase difference term 
\begin{equation}
 mc^2 \Delta \tau \approx  \frac{m}{2} \int v'^2 dt
\end{equation} 
survives in the non-relativistic limit even though it has no simple interpretation in the non-relativistic limit.

This algebraic fact (but not its origin or connection with the limiting process, etc.) has been noted earlier and discussed in a different context in ref. \cite{greenberger} where it was concluded that there are some difficulties in reconciling the Galilean transformation with quantum mechanics. Our analysis shows that this is not the case and that such a result is indeed \textit{expected} when the Galilean transformation is used correctly in quantum mechanics. The real surprise --- for which we have no simple explanation --- is the result  that the relativistic action remains form invariant even when we work to the accuracy of $\mathcal{O}(1/c^4)$.

The above result, as well as the invariance of the truncated action in \eq{quadaction}, rely crucially upon our retaining the $mc^2$ term in all the expressions. In fact, if we define a non-relativistic energy as $E_{\rm NR} \equiv mc^2 + (1/2) mv^2$, then one can again show that the phase of the wave function given by ($E_{\rm NR} t - px$) transforms to ($E'_{\rm NR} t' - p'x'$) under a Lorentz transformation, if we again retain terms only to the accuracy of $\mathcal{O}(1/c^4)$.

\section{Principle of Equivalence in Quantum Mechanics}\label{sec:qmequivalence}

We next take up issues which arise when we switch on $G$  and move from quantum mechanics towards GQM in Fig. \ref{fig:physicscube}.  One of the key questions which has been debated extensively in the literature is the validity of the principle of equivalence in quantum mechanics \cite{urev,gqm}.
In Newtonian gravity, the principle of equivalence can be introduced in many ways and it also leads to several different conclusions. We cannot, \textit{ a priori}, expect all these descriptions of the principle of equivalence (or the resulting conclusions) to be equally applicable when we proceed from the vertex QM to GQM in Fig. \ref{fig:physicscube}. We provide a brief critique of this issue in this section, essentially arguing that much of the debate can be settled by using an appropriate definition of the principle of equivalence in the quantum mechanical domain.

\subsection{The many facets of the principle of equivalence}

Consider the following two versions of the principle of equivalence: (a) The dynamics of a system in a spatially homogeneous (but possibly time dependent) gravitational field $g(t)$ is the same as the dynamics of the same system viewed in an accelerated frame with coordinates $t'=t, \ x'= x-\xi(t) $ and 
$ \ddot \xi =g$.  (b) The accelerations of different masses $m_1, m_2 ....$ in a gravitational field are independent of the masses $m_1, m_2 ...$. Therefore, any measurement based on the trajectories of a particle moving in a given gravitational field cannot be used to determine its mass. 

It is obvious that the statement (a) implies statement (b) in non-relativistic mechanics; we know that the acceleration $a$ produced in the non-inertial frame is just $a=\ddot \xi $ which is indeed independent of the mass of the particle. If the dynamics in the non-inertial frame is the same as that in a gravitational field, then we must have $a=g(t)$ in the gravitational field implying statement (b). Further, Newton's law tells us that $f=m_i a(t) = m_g g(t)$ implying equality of inertial and gravitational masses. (In this paper, the term `gravitational mass' refers to the \textit{passive gravitational mass} of the object.) Obviously, a measurement of the trajectory of the particle cannot be used to determine its mass.

A moment of thought shows that the last statement requires qualification in quantum theory.
We do not have trajectories in quantum theory but one assumes that operators like $\hat x, \hat x^2$, etc. are observables in quantum mechanics. If so, one should be able to devise a suitable experiment which measures, say, the dispersion in the position: $\sigma^2 \equiv \langle (\hat x - \langle\hat x\rangle )^2\rangle$. This quantity certainly depends on the mass of the particle when the particle moves in a gravitational field or even if the particle is free.
In fact, the case of the free particle drives home the point without clouding the issue by involving acceleration, gravity etc. The trajectory of a free particle in non-relativistic mechanics is certainly independent of its mass, and any observation related to the trajectory cannot determine its mass. But 
consider two free particles having masses $m_1$ and $m_2$ ($m_1 \neq m_2$) with identical initially prepared states in which the wave functions are Gaussian wave packets of mean zero and width $\Delta x$. If we  allow these wave packets to propagate in time, then the widths $\Delta x_1$ and $\Delta x_2$ of these wave packets at any later time $\Delta t$ will be given by:
\begin{eqnarray}
 (\Delta x_1)^2 =  (\Delta x)^2 + \dfrac{\hbar^2(\Delta t)^2 }{4 m_1^2 (\Delta x)^2}; \qquad
 (\Delta x_2)^2 =  (\Delta x)^2 + \dfrac{\hbar^2(\Delta t)^2 }{4 m_2^2 (\Delta x)^2}
\end{eqnarray} 
Hence, the width of the wave packet at any later time depends on $m$, the mass of the free particle. By measuring the width at a later time and knowing the initial width, we can deduce the mass of the particle. 
The same result holds for particles moving under a gravitational field (see e.g., \cite{tpcqg}). 
Clearly, this cannot be used to prove any violation of the principle of equivalence. 

Note that the \textit{Heisenberg equations of motion} for the quantum mechanical particle are identical in structure to the classical equations of motion and read $\ddot{\hat x}=g(t)$,  and hence, are independent of mass. Nevertheless, the dispersion in particle position will depend on the mass, in general. This shows that all the consequences which we derive from a particular version of the principle of equivalence in the classical domain may not hold in the quantum domain.

As another example illustrating the same result,
consider the context of the gravitational hydrogen atom, with the Coulomb force between the proton and the electron being replaced by the gravitational force between two particles of masses $m$ and $M$ respectively, with $m$ orbiting around $M$ and $m\ll M$. For this system, the energy levels are given by:
\begin{equation}
 E_n = -\dfrac{ G^2 M^2 m^3}{2 \hbar^2 n^2}.
\end{equation} 
 We see that the frequency of transition from one energy state to another will also be proportional to $m^3$, and hence will depend on the mass of the orbiting particle. Thus, in principle,  this frequency can be used to determine $m$. More generally, we can use the frequency of  transitions between successive energy levels of a system composed of a massive particle orbiting around a ``nucleus'', to determine the mass of the orbiting particle. This is clearly impossible in the classical theory. 
 
 Again, this experimental determination of mass should not be taken as a violation of the principle of equivalence. 
 To see what is involved, let us consider another example consisting of a particle of mass $m$, moving under the action of  a potential: 
\begin{equation}
\label{powerlawpot}
 V(x) = \alpha |x|^n
\end{equation} 
This represents a ``non-Hookean'' spring when $n\neq 2$. We will now choose the spring constant $\alpha$ such that it is proportional to the mass of the particle: $\alpha \propto m = km$. 
Classically, the mass $m$ cancels out of the equations of motion, $m\ddot x = knm|x|^{n-1}$,
showing that the acceleration is independent of the mass of the particle. (More elaborately, one can consider several masses $m_1,m_2,....$, each moving under a separate potential with the same $n$ and $k$, so that each potential scales with the corresponding mass.) Once again, classically, we cannot determine the mass by observations of the trajectory. The situation is  different in quantum mechanics.
We can  evaluate the energy levels of this ``oscillator'' in the WKB approximation using the Bohr-Sommerfeld quantization rule:
\begin{equation}
\label{bswkb}
 \int p_x dx=\int (2 m (E_N - V(x)))^{1/2} dx = N\hbar
\end{equation} 
where $N$ is the quantum number corresponding to the energy level  $E_N$, and is assumed to be  large compared to unity.
Using this integral we can determine the scaling of $E_N$ with $N$ and $m$ for different values of the exponent $n$. We find (Appendix \ref{app3}) that:
\begin{equation}
 E_N \propto N^{2n/(n+2)} \ m^{(n-2)/(n+2)}
\end{equation}
 When $n = 2$, $E_N \propto N$ which is the case of the harmonic oscillator; in this case,  $E_N$ is also independent of the mass $m$. However, when $n\neq 2$ the energy levels  $E_N$ and hence the spacing between energy levels  depend on $m$. Therefore, we can use quantum mechanical transitions between successive energy levels of any of these oscillators to determine their mass $m$! Clearly, quantum mechanics allows such a determination which is not possible in classical theory based on measurements related to the trajectory of a particle. As the above example shows, this fact has nothing to do with gravity \textit{per se} and is a general phenomenon. (In fact, even in classical mechanics, if one measures dynamical variables like momentum, energy etc., one can certainly determine the mass of a free particle.)

\subsection{Schrodinger equation in non-inertial frames and the principle of equivalence}

We believe that issues related to the determination of the inertial mass of a particle by quantum mechanical observations are red herrings to the proper formulation of the principle of equivalence in quantum mechanics. Such a formulation can be based on the statement (a) in the beginning of the last section which embodies the mathematical equivalence of dynamical equations in an accelerated frame and in a gravitational field. 
 Given the Schrodinger equation for a free particle in one frame of reference, if we transform it to any other arbitrarily accelerated frame of reference, we should obtain an equation identical in form to the Schrodinger equation for a particle in the presence of an appropriate, in general  time-dependent, \textit{gravitational field}. For example, if we transform from a frame $S = (t,x)$ to an arbitrarily accelerated frame $S' = (t, x - \xi(t))$, where $\xi(t)$ is a function of $t$, the Schrodinger equation should pick up a ``potential energy term'' $V(x) = - m \ddot{\xi}(t) x$ where $-\ddot{\xi}(t)$ is the instantaneous acceleration of $S'$ with respect to $S$. 
 
 We have already demonstrated this result earlier in \eq{schrotrans}. The only new feature is that the wave functions in the two frames are not related by a scalar transformation, but require an additional phase factor to be added. This phase factor, as we demonstrated, was required for consistency under Galilean transformations, but the situation is less clear for the case of an accelerated transformation.
 In what follows, we shall investigate this result in greater detail from different perspectives.
 
 \section{The non-relativistic limit in the presence of gravity, and the principle of equivalence}\label{sec:nringrav}
 
 In Section \ref{subsec:lorgal}, we showed that while the phase of the free particle Klein-Gordon equation is Lorentz invariant, the corresponding phase of the Schrodinger wave function is not invariant under a Galilean transformation. We also  demonstrated that the transformation to an accelerated frame in the non-relativistic context leads to the Schrodinger equation with a gravitational potential. However, to obtain this result, we needed to explicitly include a phase in the wave function in \eq{wavefntrans1}. The question arises as to how we can understand this phase in a general context when we are no longer dealing with the Lorentz transformation, and the Galilean transformation as its limiting form, but have to deal with non-inertial frames.  It is important to obtain the result in \eq{schrotrans} directly from a fully generally covariant equation, just as we obtained the results for the Galilean transformation by taking the appropriate limit of the Klein-Gordon equation. We shall now show how this can be done for a time dependent acceleration, which does not seem to have been discussed in the literature before.

 In the relativistic case, we have a quantum scalar field satisfying the free-particle Klein-Gordon equation in one frame of reference, which we call $(x,t)$; the frame $(x,t)$ being arbitrarily accelerated with time-dependent acceleration $g(t)$ with respect to an inertial coordinate system $S' = (X,T)$. In this frame (known as the \textit{generalized Rindler frame}) the metric is given by:
\begin{equation}
\label{metric1}
 ds^2 =  (1 + g(t) x/c^2) dt^2 -  dx^2
 \end{equation} 
 The explicit co-ordinate transformation \cite{TPgr} between $S$ and the inertial frame $S'$ is given by:
 \begin{equation}
 X =  c\int{dt \ (1 +  g(t) x/c^2)\sinh \chi(t) };\qquad
 T =  \int{dt \ (1 +  g(t)x/c^2)\cosh \chi(t)}
\end{equation}  
where $\chi(t)$ is related to $g(t)$ by:
 \begin{equation}
  g(t) = c\frac{d \chi}{dt}
 \end{equation} 
 The  generally covariant Klein-Gordon equation for a scalar field $\Phi(x,t)$ in this frame is given by:
 \begin{equation}
 \frac{1}{\sqrt{-g}} \partial_i ( \sqrt{-g} g^{ik} \partial_k \Phi) = - m^2 \Phi
\end{equation} 
 Hereafter, unless mentioned otherwise, we use units in which $c = \hbar = 1$.
Using the form of the metric as given in \eq{metric1}, we find:
\begin{equation}\label{kgarbaccn}
 -\dfrac{1}{(1 + g(t)x)^2}\frac{\partial^2\Phi }{\partial t^2} + x \frac{d g}{d t}\frac{\partial \Phi}{\partial t}\dfrac{1}{(1 + g(t)x)^3} + \frac{\partial^2 \Phi}{\partial x^2} + \frac{\partial \Phi}{\partial x}\dfrac{g}{(1 + g(t)x)} = m^2 \Phi
\end{equation}  
We now substitute $\Phi(x,t) = \psi(x,t) e^{-imt}$ and note the following:
\begin{equation}
 \frac{\partial \Phi}{\partial t} = -ime^{-imt}\psi(x,t) + e^{-imt}\frac{\partial\psi}{\partial t}; \quad \frac{\partial \Phi}{\partial x} = e^{-imt} \frac{\partial \psi}{\partial x} 
\end{equation} 
and
 \begin{equation}
  \frac{\partial^2 \Phi}{\partial t^2} = -m^2 e^{-imt}\psi(x,t) -2 i m e^{-imt}\frac{\partial\psi}{\partial t} + e^{-imt}\frac{\partial^2 \psi}{\partial t^2}; 
 \end{equation} 
 
 \begin{equation}
  \frac{\partial^2 \Phi}{\partial x^2} =  e^{-imt}\frac{\partial^2 \psi}{\partial x^2}; 
 \end{equation} 
 Substituting the above relations into \eq{kgarbaccn}, we find (putting back the $c$- and $\hbar$- factors):
 \begin{eqnarray}
-\left[\left(1 + \frac{g(t) x}{c^2}\right)\right] \left[-\frac{m^2 c^2 \psi}{\hbar^2} -  \frac{2i m}{\hbar} \frac{\partial \psi}{\partial t} + \frac{1}{c^2}\frac{\partial^2 \psi}{\partial t^2}\right]   &-& \nonumber\\ \left(\frac{imc^2\psi}{\hbar} -  \frac{\partial \psi}{\partial t}\right)\frac{x}{c^4}\frac{dg}{dt} + \left(1 + \frac{g(t) x}{c^2}\right)^3 \frac{\partial^2 \psi}{\partial x^2} &+& \nonumber \\ \left(1 + \frac{g(t) x}{c^2}\right)^2 \frac{\partial \psi}{\partial x}\frac{g}{c^2} = \frac{m^2 c^2}{\hbar^2} \left(1 + \frac{g(t) x}{c^2}\right)^3\psi
 \end{eqnarray} 
 Retaining terms to the lowest order (upto but excluding order $gx/c^2$), we get on simplification:
 \begin{equation}
  i \hbar\frac{\partial\psi}{\partial t} =  -\frac{\hbar^2}{2m}\frac{\partial^2 \psi}{\partial x^2} + mg(t)x\psi
  \label{gravschrodeqn}
 \end{equation} 
 which is identical to the Schrodinger equation for a particle of mass $m$ in an accelerated frame of reference moving with acceleration $-g(t)$ or equivalently, in a time-dependent gravitational field of strength $g(t)$.
 Hence, we see that the Klein-Gordon equation reduces in the appropriate limit to the Schrodinger equation, with the term $mg(t)x$ indicating the accelerated nature of the frame. 
 
 Notice that we have excluded the terms of $1/c^2$ order. This is consistent with our earlier comments on the ``accuracy'' of the Galilean transformation. For the Galilean transformation and the non-relativistic limit to be valid, one must keep only those terms upto order $(1/c)$ and not beyond.
 
 In fact, the phase $f$ acquired by the non-relativistic wave function acquires a direct physical meaning when we consider the Schrodinger equation as a limit of the Klein-Gordon equation.  We can  transform the free particle solution to the Klein-Gordon equation in the inertial frame, $\Phi(T,X)$, as a scalar to the non-inertial frame, thus obtaining $\Phi(t,x)$. The non-relativistic limits of $\Phi(T,X)$ and $\Phi(t,x)$ will differ by a phase term $mc^2(t-T)$, which, in the appropriate limit, will give the correct phase dependence arrived at in \eq{wavefntrans} when we consider the effect of gravitational time dilation. In the presence of a gravitational potential $\phi$, the proper time lapse $dT$ of a co-moving clock is related to the coordinate time lapse $dt$ by  
 \begin{equation}
  ds^2 = c^2dT^2=c^2 \left(1 + \frac{2 \phi}{c^2} \right) dt^2 - dx^2 
       = c^2 dt^2 \left[\left(1 + \frac{2 \phi}{c^2} \right) - \frac{V^2}{c^2}\right] 
 \end{equation} 
 so that, when $V=\dot\xi, \phi=x\ddot\xi$, we get 
 \begin{eqnarray}
  mc^2(t-T)&=&-mc^2\left[ 
  \int dt \left( 
  1- \frac{\dot\xi^2}{c^2} +\frac{2x\ddot\xi}{c^2}
          \right)^{1/2}
	  -t    \right]\nonumber\\
&\approx&-m 
  \int dt \left( 
  - \frac{\dot\xi^2}{2} +x\ddot\xi
          \right)
= -mx\dot\xi+\frac{1}{2}m\int dt \dot\xi^2 
 \end{eqnarray} 
which is precisely the phase $f$ found in \eq{fform}! Once again we see that a result in non-relativistic quantum mechanics acquires a simple interpretation when we treat it as a limit of relativistic theory, thanks to the factor $mc^2(t-T)$ in the phase. The result also shows that in the instantaneous rest frame of the particle, the phase of the wave function evolves as $mc^2 d\tau$, where $\tau$ is the proper time shown by the co-moving clock, thereby again validating the principle of equivalence in quantum mechanics.

We conclude this section with a comment related to the structure of the Schrodinger equation in 
\eq{gravschrodeqn} and its relationship to quantum field theory in the Rindler frame. For a general $\xi(t)$, the potential in \eq{gravschrodeqn} depends explicitly on time, and hence it does not allow stationary state solutions. The exception occurs in the case of a uniformly accelerated frame for which $\ddot\xi=g$ is a constant. In that case, \eq{gravschrodeqn} possesses stationary state solutions with the energy eigenfunctions being Airy functions. A similar situation arises in the fully relativistic context. For a time dependent $g(t)$, the metric in \eq{metric1} is not stationary and the Klein-Gordon equation does not possess mode functions with an $\exp(-iEt)$ dependence. The exception is the case of constant $g$, leading to the Rindler metric, when such stationary solutions exist. (In fact, these solutions, given in terms of Hankel functions, reduce to the corresponding Airy functions in the $c\to\infty$ limit, as they should.). We know, however, that the positive frequency mode functions in the Rindler frame are a superposition of positive and negative frequency mode functions in the inertial frame, leading to \cite{unruhdavies} the well-known Davies-Unruh temperature $T=(\hbar/c)(g/2\pi)$. When one takes the $c\to\infty$ limit, this effect vanishes at the lowest order (which is independent of $c$) but leads to a nontrivial result at the next order. We hope to describe this and related features in a separate publication \cite{hptpinprog}.

 \section{Propagators in relativistic and non-relativistic quantum theories}
 \label{sec:relnonrelprop}
 
 In the previous sections we dealt with the Klein-Gordon equation and the Schrodinger equation without worrying about the physical interpretation of their solutions in the relativistic and non-relativistic contexts. It is, however, well known that while the non-relativistic wave function has a standard interpretation, as the probability amplitude in a single particle theory, the solution to the Klein-Gordon equation cannot be interpreted as a probability amplitude for the relativistic particle. In fact, consistent interpretation of quantum field theory requires us to recognize the existence of anti-particles and treating the solutions of relativistic wave equations as \textit{operator-valued} entities. This requires a closer scrutiny of  the limits which we studied in the previous sections, which we shall now turn our attention to.
 
 The solution $\phi(x^i)$ of a Klein-Gordon equation treated as an operator, can be expanded using creation ($a^\dagger_{\bf k}$) and annihilation ($a_{\bf k}$) operators in the second quantized formalism:
 \begin{equation}
  \phi(x^i) = \sum_{\bf k} (a_{\bf k} f_{\bf k} + a^\dagger_{\bf k}f^*_{\bf k})
 \end{equation} 
 where $f_{\bf k}(x^i)$ is a c-number solution to the Klein-Gordon equation labelled by the momentum ${\bf k}$. The transition element $\langle 0|\phi|1_{\bf k}\rangle = f_{\bf k}$ between the standard vacuum state $|0\rangle$ and the one-particle state $|1_{\bf k}\rangle$ with momentum $\mathbf{k}$ is what we have taken to be the relativistic analog of a wave function in the previous sections. In the $c\to \infty$ limit, the combination $ \psi=f_{\bf k} e^{imc^2t} $
 will reduce to the Schrodinger wave function for the free particle.  In a similar way, one can think of the state $\phi(x) |0\rangle$ as representing a state with a particle at position $x^i$ and the quantity 
 \begin{equation}
G_F(x,y) = \langle 0| T[\phi(x)\phi(y)]|0\rangle                                                       \end{equation} 
 as the amplitude for a particle to propagate from the event $y$ to event $x$ with the time ordering operator $T$ to take care of the distinction between particles and anti-particles.
 
 Relativistic quantum field theory makes use of the (Feynman) propagator $G_F$ extensively and, in fact, all the standard results of interacting field theory can be obtained using this propagator. It is, of course, possible to introduce a similar propagator in non-relativistic quantum mechanics, which can also be interpreted as the amplitude for a particle to propagate from one event to another, even though it is not as extensively used as the propagator in field theory. Given the fact that the solutions to the Klein-Gordon equation have no direct interpretation as probability amplitudes in quantum field theory, it is better to study the non-relativistic limit of quantum field theory in terms of the two propagators. In the present section, we shall consider different aspects of this issue. 
 
 \subsection{Path integral description of relativistic and non-relativistic particles}
 
 In non-relativistic quantum mechanics, one can determine the wave function at $t=t_2$ from the wave function at $t=t_1$, where $t_2\ge t_1$, by a suitable integral kernel $K$  via the equation:
 \begin{equation}
 \psi(\mathbf{x}_2,t_2) = \int K(\mathbf{x}_2, t_2; \mathbf{x}_1, t_1) \ \psi(\mathbf{x}_1,t_1) d\mathbf{x}_1
 \label{five}
\end{equation} 
This propagation kernel can be expressed formally as a sum over paths $\mathbf{x}(t)$ of an amplitude $\exp(i\mathcal{A}[\mathbf{x}(t)])$ where $\mathcal{A}[\mathbf{x}(t)]$ is the action for the path satisfying the appropriate boundary conditions at $t=(t_1,t_2)$. For a free particle of mass $m_0$, the kernel is given by:
 \begin{equation}
 K = \sum_{\mathbf{x}(t)} \ \exp\left( \frac{i}{\hbar} \int_{t_1}^{t_2} \frac{1}{2} m_0 {\bf |\dot x|}^2 \ dt\right)
\end{equation}

 The sum over paths can be rigorously defined by a time-slicing procedure which assumes that at any intermediate time $t$, with $t_1\le t\le t_2$, the particle has a unique position.
 If the action is a quadratic functional, it can be shown that (see e.g.,\cite{feynmanhibbs}) the propagation kernel has the form:
 \begin{equation}
 K(\mathbf{x}_2, t_2; \mathbf{x}_1, t_1) = N(t_1,t_2)\exp {\frac{i}{\hbar}}\mathcal{A}_c(t_2,\mathbf{x}_2;t_1,\mathbf{x}_1)
\end{equation}
where $\mathcal{A}_c$ is the action for the classical path.
 In particular, the kernel for a free particle is given by:
\begin{eqnarray}
 K(\mathbf{x}_2, t_2; \mathbf{x}_1, t_1) &=& \left( \frac{m_0}{2\pi i \hbar (t_2-t_1)}\right)^{D/2} \, 
 \exp\left[\frac{im_0(\mathbf{x}_2-\mathbf{x}_1)^2}{2\hbar(t_2-t_1)}\right]
 \,\theta(t_2-t_1)\nonumber\\
 &\equiv&
  \theta(t_2-t_1) F(\mathbf{x}_2, t_2; \mathbf{x}_1, t_1)
\end{eqnarray} 
where $D$ is the number of spatial dimensions.
 (The $\theta$ function is just conventional;  purely algebraically, \eq{five}
 holds even for $t_2<t_1$.) It is also often convenient to define the propagator in \textit{the  energy space} by the definition:
 \begin{equation}
 G(E, \mathbf{x_2},\mathbf{x_1})= \int_{-\infty}^\infty dT e^{iET} K(T,\mathbf{x_2},\mathbf{x_1}) =\int_0^\infty dT e^{iET} F(T,\mathbf{x_2},\mathbf{x_1})
\end{equation}
 where $T=t_2-t_1$. This holds for any time-independent action functional;  for a free particle, we have the result: 
 \begin{equation}
 G(E, \mathbf{x_2},\mathbf{x_1})=\left( \frac{m_0}{2\pi i }\right)\,\frac{1}{|\mathbf{x_2}-\mathbf{x_1}|}\, \exp[-i\sqrt{2m_0E|\mathbf{x_2}-\mathbf{x_1}|^2]}
\end{equation} 
 From \eq{five}, it also follows that: 
 \begin{equation}
 F(t_3, \mathbf{x}_3; t_1,\mathbf{x}_1) = \int d\mathbf{x}_2 F(t_3,\mathbf{x}_3; t_2,\mathbf{x}_2) \ F(t_2,\mathbf{x}_2; t_1,\mathbf{x}_1)
 \label{ten}
\end{equation} 
 which we will call the \textit{transitivity property} of the kernel. The same result, of course, holds for $K$ as well but we explicitly express it in the above form (using $F$) to stress that --- purely algebraically --- this result is valid even without any time ordering.
 
 We do \textit{not} expect something analogous to \eq{five} to hold in the relativistic case. A  mathematical reason for this fact is simple: The relativistic mode function satisfies a second order differential equation in time, unlike its nonrelativistic counterpart which satisfies the Schrodinger equation that is first order in time. Hence the mode function at time $t_2$ cannot be determined given \textit{only} the mode function at an earlier time, unlike in the case of \eq{five}. In fact, if the mode function $\phi$ and its first time derivative are known at some initial time $t_1$, then, at some other time $t_2$, we have:
\begin{equation}
\phi(\textbf{x}_2, t_2) = \int d^3\textbf{ x}_1 \left[G (\textbf{x}_2, t_2; \mathbf{x}_1, t_1) \frac{\partial  \phi(\mathbf{x}_1, t_1)}{\partial t_1} - \frac{\partial G (\mathbf{x}_2, t_2;\mathbf{ x}_1, t_1)}{\partial t_1}  \phi(\mathbf{x}_1, t_1)\right]
\end{equation}
where $G (\mathbf{x}_2, t_2;\mathbf{ x}_1, t_1)$ is the retarded Green's function. It is obvious that both $\phi$ and its first derivative are required to be known at the initial time $t_1$ in order to determine $\phi$ at the later time $t_2$.

 Nevertheless, given the fact that we do have a valid action functional for a relativistic particle, one can formally define a kernel using the path integral as
 \begin{equation}
K (x^i;y^i) = \sum\limits_{all x^i(s)}~~ \exp~[i~\mathcal{A}[x(s)]/\hbar]
\label{eleven}
\end{equation} 
 where 
 \begin{equation}
\mathcal{A}= -m_0 \int \sqrt{1-\dot{\mathbf{x}}^2}\ dt
=-m_0\int_0^1d\eta \left(\frac{dx^i}{d\eta}\frac{dx_i}{d\eta}\right)^{1/2}
 \label{twelve}
\end{equation}
The second form of action expresses the trajectory $x^i(\eta)$ in terms of a parameter $\eta$ chosen such that $x^i(0)=y^i,x^i(1)=x^i$ are the boundary values.
  Unfortunately, this action is not quadratic in the dynamical variables and hence the standard procedure for evaluating the path integral fails. 
 
 It is, however, possible to evaluate the path integral in \eq{eleven} \textit{exactly} by at least two other  methods. The first involves utilizing the fact that the integral in \eq{twelve} is just the proper length of a given path connecting the end points. We can implement the sum over paths in a Euclidean lattice and take an appropriate limit to the continuum at the end, thereby evaluating \eq{eleven}. This is worked out in detail in reference \cite{tpfop}. 
 
 A second procedure is more heuristic, but physically transparent, and is based on the fact that even in non-relativistic quantum mechanics, there exists an action functional, called the Jacobi action, which has a structure similar to the one in \eq{twelve}
 involving a square root. We shall briefly indicate how this procedure can be used to explicitly evaluate the path integral in \eq{eleven}. We shall essentially prove that  the following relation holds: 
 \begin{eqnarray}
&& \sum_{all\,{\bf x}}\exp {\frac{i}{\hbar}}\int^1_0 ds
\sqrt{2m_0(E-V)|{\bf \dot x}|^2}\nonumber\\
&&\qquad \qquad=\int\limits^{\infty}_0 dTe^{{\frac{i}{\hbar}}ET}
\left\{\sum_{x(t)} \exp {\frac{i}{\hbar}}\int^T_0 dt \left[\frac{1}{2} m_0{\mathbf{\dot x}}^2 - V\right]\right\}
\label{nineteen}
\end{eqnarray}
(On either side of the equation, the overdot denotes differentiation with respect to the integration variable used.)
 The expression on the left hand side is a path integral defined using the Jacobi action for a particle propagating from one spatial point to another with energy $E$ in an external potential $V$. The right hand side involves the Fourier transform of the usual path integral (given in the curly brackets) which we know how to evaluate. This relation allows us to give meaning to path integrals involving re-parameterization invariant action functionals which have square-roots in them. The proof of \eq{nineteen} is given in Appendix \ref{app4}.
 
 We can evaluate the path integral for the relativistic particle involving a square root in the action using the result in \eq{nineteen} by making the following substitutions in \eq{nineteen}:
 (i) $E \to m$, (ii) $ m_0 \to m/2$, (iii) $V=0$, and a change of sign on the integration variable.  Thus, \eq{nineteen} becomes:
 \begin{equation}
 \sum\limits_{paths} \exp - {i \over \hbar} \int\limits_0^1 ds ~  m  \sqrt {\dot x_i \dot x^i}  = \int\limits_0^{\infty} d \tau ~e^{-im\tau/\hbar} \left\{ \sum_{x^i}~ \exp -
{i \over \hbar} ~{m \over 4} \int\limits_0^{\tau} {\dot x^i \dot x_i \ ds}\right\}
\label{twenty}
\end{equation} 
On transforming the integration variable on the left-hand side of the above equation from $s$ to $t = Ts$, we find that it becomes:
\begin{equation}
 \sum\limits_{paths} \exp - {i m \over \hbar} \int\limits_0^T dt  \sqrt {\frac{dx_i}{dt}\frac{dx^i}{dt}}  = \sum\limits_{paths} \exp - {i m \over \hbar} \int\limits_0^T dt \sqrt{1 - \dot{\mathbf{x}}^2}
\end{equation} 
which is exactly the relativistic propagator with the square-root action that we were interested in evaluating (\eq{eleven}).

 The sum over paths within the curly brackets on the right-hand side of \eq{twenty} can be evaluated by the standard procedure for a free particle which allows the relativistic kernel to be expressed as the integral
 \begin{equation}
K =  \frac{m}{16 \pi^2}\int_0^\infty\frac{d s}{ s^2} \, e^{-im^2 s}\ e^{-il^2/4s} = (i m) G_F; \qquad l^2 = (x-y)^2
 \label{twentyone}
\end{equation} 
 where we have done a rescaling from $\tau$ to   $s\equiv (\tau/m)$ to express it in the standard form. The $G_F$ is the standard Feynman propagator 
 in Schwinger's proper time representation, which differs from the kernel by an unimportant constant. (Dimensionally, this constant is to be expected, because the Feynman propagator has dimensions of square of inverse length, while the kernel which we have evaluated has the dimensions of cube of inverse length). The way we have obtained the relativistic kernel shows that it is actually the result of evaluating a path integral, with the action for a relativistic particle --- as to be expected. Hereafter, we shall work with the Feynman propagator $G_F$, as it is the usual quantity of interest in quantum field theory.
 
 As usual, the factor $\exp{-im^2s}$ is not strongly convergent in the upper limit of the integration. This difficulty can be tackled by interpreting $m^2$ as $(m^2-i\epsilon)$ or by rotating the integration contour to the imaginary axis by going over to the Euclidean sector. The Euclidean propagator is given by:

 \begin{equation}
\label{feynpropeucl}
 G_F=\frac{1}{16\pi^2} \int^\infty_0 \frac{d\tau_E}{\tau^2_E} \, e^{-m^2 \tau_E} \, e^{- x_E^2/4\tau_E} = \frac{m}{4 \pi^2} \frac{1}{\sqrt{x_E^2}} K_1 (m \sqrt{x_E^2})
\end{equation}
where $K_1$ is the modified Bessel function of the second kind.

 The above expression can indeed be interpreted as the amplitude for a relativistic particle to propagate from one event $y^i$ to another event $x^i$, and hence is equal to  $\langle 0|T[\phi(x)\phi(y)]|0\rangle$. But it does not propagate any ``wave function'' in the relativistic theory for reasons explained earlier. We, however, would expect this propagator to reduce to the standard non-relativistic kernel in the $c\to \infty$ limit. We shall now describe how this comes about. 
 
 \subsection{Non-relativistic limit of Feynman propagator in inertial  frame}
 
 Let us first consider the non-relativistic ($c\to\infty$) limit of \eq{twentyone}. It is easy to see that, in this limit, the integral in \eq{twentyone} can be evaluated in the saddle point approximation. (In this section, we set $\hbar = 1$.)
 To avoid mathematical ambiguities, it is convenient to work in the Euclidean space and  start with \eq{feynpropeucl}:
 \begin{equation}
G_F=\frac{m}{16\pi^2} \int^\infty_0 \frac{d\tau_E}{\tau^2_E} \, e^{-m \tau_E} \, e^{-m x_E^2/4\tau_E}
\end{equation} 
where we have rescaled $\tau_E$ in \eq{feynpropeucl} to $\tau_E/m$ so that $\tau_E$ has the dimensions of length (or time).
 The stationary phase of this integral occurs at 
 $\tau_E^2 = (x_E^2/4 )$, where $x_E^2 = t_E^2 + |\mathbf{x}|^2$,
 so that the integral evaluates to
\begin{eqnarray}
\label{Gsaddleptfree}
 G_F &=& N\frac{m}{4\pi^2 x_E^2} \exp -mc\sqrt{x_E^2} \nonumber\\
 &=& N \frac{m}{4\pi^2 x_E^2} \exp -mc^2 t_E \sqrt{1+ \frac{|\mathbf{x}|^2}{c^2 t_E^2}}
\end{eqnarray} 
where we have put back the $c-$factors in the phase.
 The prefactor is given by: 
 \begin{equation}
  N = \left(\frac{2\pi}{f^{''}(\tau_0)}\right)^{1/2} = \left(\frac{\pi \sqrt{x_E^2}}{2 m}\right)^{1/2}
 \end{equation} 
 where $\tau_0 = \sqrt{x_E^2}/2$ is the saddle point of the integrand and $f(\tau_E) = - m\tau_E - mx_E^2/4\tau_E$ is the argument of the exponential.

We now take the $c \to \infty$ limit of the above expression. Noting that, in this limit, 
 $\tau_E^2 = x_E^2/4 c^2 \simeq t_E^2/4$ where $x_E^2 = c^2 t_E^2 + |\mathbf{x}|^2$, we get (relabeling $G_F$ to $K$, the kernel):
\begin{equation}
 K \simeq  \left(\frac{1}{2m}\right)\left( \frac{m}{2\pi t_E}\right)^{3/2} \ \exp (-mc^2t_E) \  \exp (-m|\mathbf{x}|^2/(2t_E))
\end{equation}  
   Finally, going back to Lorentzian time 
 with $t_E = it$, we obtain:
\begin{equation}
 K \simeq \left(\frac{1}{2m}\right)\left(\frac{m}{2 \pi  i  t}\right)^{3/2}  \ \exp (-i mc^2t) \  \exp (i m|\mathbf{x}|^2/(2 t))
 \label{lorenttime}
\end{equation} 
which matches exactly with the non-relativistic path-integral kernel for a free particle in three dimensions, apart from two standard factors:  (i) The factor $\exp ( - i mc^2t)$ is the standard extra term in the action functional which arises on taking the non-relativistic limit of the fully relativistic action functional. This is exactly the phase term which we factored out in the wave function, while going from the Klein-Gordon equation to the Schrodinger equation for an accelerated particle, previously in Section \ref{sec:nringrav}. (ii) The factor $1/2m$ arises due to the difference in normalization conventions for the relativistic and non-relativistic mode functions. In non-relativistic quantum mechanics, the free-particle wave functions are normalized with the factor $(2 \pi) ^{-3/2}$, while in quantum field theory, the mode functions are normalized with the factor $(2 \pi) ^{-3/2} (2 E_{\mathbf{p}})^{-1/2}$. Since the Feynman propagator is bilinear in the mode functions, this leads to an extra factor $(1/ 2 E_{\mathbf{p}})$  in the Feynman propagator as compared to the path-integral kernel. The factor $(1/ 2 E_{\mathbf{p}})$ in the non-relativistic limit reduces to just $(1/2m)$.

By the same procedure, we can also compute the first-order relativistic correction to the path integral. To do this, we retain one more order in $(1/c)$ in the expansion of 
 $(1+ |\mathbf{x}|^2/c^2 t_E^2)^{1/2}$ in \eq{Gsaddleptfree} to obtain: 
 \begin{equation}\label{kcorrection}
  K' = K \left(1 - \frac{3 |\mathbf{x}|^2}{4 c^2 t_E^2}  + \frac{m|\mathbf{x}|^4}{8 c^2 t_E^3} + \mathcal{O}\left(\frac{1}{c^4}\right)\right)
 \end{equation} 
  This gives the correction to the non-relativistic propagator to one higher order in $1/c$.
  We hope to discuss the physics from these corrections in a future work.

 The above analysis shows that the Feynman propagator of the field theory does reduce to the correct free particle propagator in non-relativistic quantum mechanics in spite of all the interpretational differences between the two theories.
 This is related to the following facts.
 
 (i) The time component of the conserved  probability current for the Klein-Gordon equation reduces to $|\psi|^2$ in the appropriate limit, allowing for the standard probabilistic interpretation.  Since, in this limit, the Klein-Gordon equation becomes effectively first order in time, we find that \eq{five} also holds to the correct order of approximation. More formally, the second order propagation equation for the mode functions of the Klein-Gordon equation does reduce to \eq{five} (with $G_F$ reducing to $K$) in the appropriate limit, making everything consistent.
 
 (ii) In the non-relativistic path integral defining the propagator $K$, we sum over paths which always go forward in time. However, in the expression in the curly bracket in \eq{twenty} which gives the corresponding path integral in four dimensions, particles go forward \textit{in proper time} but their paths may go forward or backward in coordinate time. On the left hand side of \eq{twenty}, this will correspond to paths with $v>c$ for which the amplitude is exponentially suppressed. All these effects vanish in the $c\to \infty $ limit, thus allowing a single particle interpretation. 
 
 (iii) The situation is very similar to that of paraxial propagation in optics, which allows us to approximate the second-order wave equation into one which involves only first derivatives with respect to a given coordinate, viz., the coordinate along which light is propagating.  
The mathematics is identical in both the cases \cite{tpparaxial}. In the case  involving a beam of light (of frequency $\omega$) traveling along the $z-$axis from a slit located at $(z_1,\mathbf{x}_1^\perp)$, say, to a point on a  screen with coordinates $(z_2,\mathbf{x}_2^\perp)$, with $(z_2-z_1)\gg |\mathbf{x}_2^\perp-\mathbf{x}_1^\perp|$ (`paraxial propagation'), the relevant physics is contained in the phase difference acquired by the beam corresponding to a path difference $\Delta l$:
\begin{equation}
\left(\frac{\omega}{c}\right)\Delta l\equiv \frac{\omega}{c}\left\{
\left[(z_2-z_1)^2+(\mathbf{x}_2^\perp-\mathbf{x}_1^\perp)^2\right]^{1/2}-(z_2-z_1)
\right\}
\approx \frac{\omega}{c}\frac{(\mathbf{x}_2^\perp-\mathbf{x}_1^\perp)^2}{2(z_2-z_1)}
\label{optics}
\end{equation}
 For a relativistic particle of mass $m$, the corresponding phase is related to the proper time lapse $\Delta\tau$ and is given by:
\begin{equation}
\frac{mc^2}{\hbar} \Delta \tau\equiv \frac{mc}{\hbar}\left\{
[c^2(t_2-t_1)^2-(\mathbf{x}_2-\mathbf{x}_1)^2]^{1/2}-c(t_2-t_1)
\right\}
\label{relpart}
\end{equation}
which reduces, in the $c\to\infty$ limit (which implies the paraxial propagation condition: $c(t_2-t_1)\gg |(\mathbf{x}_2-\mathbf{x}_1)|$ along the time axis), to:
\begin{equation}
\frac{mc^2}{\hbar}\Delta \tau
\approx \frac{m}{\hbar}\frac{(\mathbf{x}_2-\mathbf{x}_1)^2}{2(t_2-t_1)}
\end{equation}
which is precisely the phase of the quantum mechanical kernel for the nonrelativistic particle, arising again from the rest mass energy term in the appropriate limit.
Just as in paraxial optics we retain the propagation of light along the forward $z$ direction (from slit to screen, say), ignoring backward propagation, in arriving at nonrelativistic quantum mechanics from field theory, we retain the propagation of the particle forward in the time axis, ignoring backward propagation. In both the contexts, the approximation effectively reduces the degree of the relevant differential equation from two to one in the appropriate independent variable (which is $z$ in  optics and $t$ in quantum theory).
That is, in arriving at non-relativistic quantum mechanics from quantum field theory, one considers paraxial propagation of the Klein-Gordon modes with the axis now being along the time direction.

 \subsection{Non-relativistic limit of the Feynman propagator in the Rindler  frame}
 
  We have already seen in Section \ref{sec:nringrav} that, at the level of differential equations, the relevant limits work out correctly: 
 The $c\to \infty$ limit of the Klein-Gordon equation  in a homogeneous, time dependent, gravitational field gives us the correct Schrodinger equation. Further, except for a phase whose origin we can understand, the same Schrodinger equation is obtained in a non-inertial frame mimicking the same gravitational field. This establishes the principle of equivalence in the context of quantum mechanics at the level of the field equations.
 
 But, as mentioned earlier, there is a significant difference in the interpretation of the solutions of the Klein-Gordon and Schrodinger equations in the two contexts. Therefore, it is not clear whether the formal equivalence of the two differential equations in the appropriate limit ensures the validity of the principle of equivalence. A more rigorous way of establishing this result would be to work at the level of propagators, and show that the Feynman Green's function in a non-inertial frame reduces in the $c\to\infty$ limit to the non-relativistic kernel for the appropriate gravitational field. We shall now describe how this comes about in the case of a constant acceleration $g(t)=g$ in the negative $x-$direction, or equivalently, a constant gravitational field $g$ in the $+x$ direction. The non-relativistic kernel for this case is given by (\cite{feynmanhibbs}):
\begin{equation}
\label{unigker}
 K(x_b, T_b; x_a, 0) = \left(\frac{m}{2 \pi i \hbar T_b} \right) \exp \frac{i m }{2 \hbar T_b }  \left((x_b - x_a)^2 - \frac{1}{12}g^2 T_b^4 - g T_b^2( x_b + x_a)\right)
\end{equation} 
 
 The transformation from an inertial frame $(T,X)$ to an accelerated frame $(t,x)$ is given by 
 \begin{equation}
 gT =  (1+gx)\sinh(gt); \quad  1+gX= (1+gx) \cosh(gt) 
 \label{twentyseven}
\end{equation} 
 The Feynman propagator in this uniformly accelerated  frame can be obtained from the propagator in the inertial frame, given by 
 \begin{equation}
  G_F = - i \int\frac{ds}{(4\pi)^2 s^2} \, e^{-im^2s}\ e^{-il^2/4s}
\end{equation} 
 by expressing the line interval $l^2$ in terms of the Rindler coordinates. (This result holds because of the bi-scalar nature of the Feynman propagator.) Elementary algebra using \eq{twentyseven} gives the line interval between the two events $(t_a, x_a)$ and $(t_b,x_b)$ to be 
 \begin{equation}
 l^2 = - \frac{1}{g^2} \left[ \left( 1+\frac{g x_b}{c^2}\right)^2 + \left( 1+\frac{g x_a}{c^2}\right)^2 -2 \left( 1+\frac{g x_b}{c^2}\right)\left( 1+\frac{g x_a}{c^2}\right)\cosh g(t_b -t_a)\right]
\end{equation} 
 so that the propagator becomes
 \begin{eqnarray}
 \label{schwingerprop}
   G_F(x_b, t_b; x_a, t_a) &=&-\int_0^{\infty}\frac{ds}{16\pi^2s^2} \exp \left[\frac{i}{4g^2 s}\left( \left(1 + \frac{g x_b}{c^2}\right)^2 +\left(1 + \frac{g x_a}{c^2}\right)^2  \right. \right. \nonumber \\
   && \left. \left. - 2 \left(1 + \frac{g x_b}{c^2}\right)\left(1 + \frac{g x_a}{c^2}\right) \cosh g(t_b - t_a) \right) - im^2 s\right]
 \end{eqnarray} 
 The  above integral can be again approximated by the method of stationary phase.
 The phase of the integrand becomes stationary for the value of $s$ given by:
 \begin{eqnarray}
  s &=& \frac{i}{2 g m} \left[ \left(1 + \frac{g x_b}{c^2}\right)^2 +\left(1 + \frac{g x_a}{c^2}\right)^2  \right.  \nonumber \\
   &&  \left. - 2 \left(1 + \frac{g x_b}{c^2}\right)\left(1 + \frac{g x_a}{c^2}\right) \cosh g(t_b - t_a)\right]^{1/2}
 \end{eqnarray} 
 Substituting for  $s$ in \eq{schwingerprop}, we get:
  \begin{eqnarray}
G_F &=& N \exp \frac{mc^3}{g\hbar}\left[\left( \left(1 + \frac{g x_b}{c^2}\right)^2 +\left(1 + \frac{g x_a}{c^2}\right)^2  \right. \right. \nonumber \\
   && \left. \left. - 2 \left(1 + \frac{g x_b}{c^2}\right)\left(1 + \frac{g x_a}{c^2}\right) \cosh g(t_b - t_a) \right)\right]^{1/2}
  \end{eqnarray} 
 where $N$ is the standard normalization factor arising from the saddle-point approximation. 
 
 We will now obtain the non-relativistic $(c \rightarrow \infty)$ limit of the above expression by expanding the exponent in a power series retaining terms upto order $(1/c^4)$ inside the radical.  This gives:
 \begin{eqnarray}
&& \frac{mc^3}{\hbar}  \left[\frac{2}{g^2} + \frac{2x_b}{c^2 g} + \frac{x_b^2}{c^4} + \frac{2 x_a}{c^2 g} +\frac{x_a^2}{c^4} - \frac{2}{g^2}\left(1 + \frac{g x_b}{c^2} +\frac{g x_a}{c^2} + \frac{g^2 x_b x_a}{c^4} \right) \right. \nonumber \\
 && \left. \left[1 + \frac{g^2 (t_b - t_a )^2}{2 c^2} + \frac{g^4 (t_b - t_a )^4}{24 c^4} + \mathcal{O} \left(\frac{1}{c^6}\right) \right]\right]^{1/2}
 \end{eqnarray} 
  On simplification, ignoring all terms of order $1/c^6$ and higher inside the bracket, we obtain:
 \begin{eqnarray}
&& -\frac{imc^2(t_b - t_a )}{\hbar}  \left[1 - \frac{(x_b - x_a )^2}{2 c^2 (t_b - t_a )} + \frac{g(x_b + x_a )}{2 c^2} + \frac{g^2 (t_b - t_a )^2}{24 c^2}\right]
 \end{eqnarray}  
  which simplifies to: (with $(t_b - t_a ) \equiv T$)
 \begin{eqnarray}
&& -\frac{imc^2T}{\hbar} +  \frac{i m (x_b - x_a )^2}{2 \hbar T} - \frac{img(x_b + x_a )T}{2 \hbar} - \frac{i mg^2T^3}{24 \hbar}
 \end{eqnarray}   
 which, apart from the term  $-imc^2T/\hbar$, is identical to the phase of the standard, non-relativistic Feynman path integral for a particle of mass $m$, moving with a constant acceleration $g$ in the negative $x-$direction, or equivalently, situated in a uniform gravitational field of strength $g$ in the positive $x-$direction, as given in \eq{unigker}; see also \eq{propuniaccn} in Appendix \ref{app5} with $\ddot \xi$ replaced by $-g$.  The normalization factor $N$  also works out correctly just as in the case of the free particle. Hence, we have shown that  the relativistic Feynman propagator in the Rindler frame reduces, in the appropriate limit, to the corresponding non-relativistic Feynman path integral for the case of a particle moving with uniform acceleration.

\subsection{Non-relativistic limit of Feynman propagator in a weak gravitational field}

We now generalize the above result and formally show how  the limiting form of the non-relativistic propagator  can be obtained for \textit{any} weak Newtonian gravitational field, starting from a fully relativistic description.  Such a weak field is described by a metric of the form
\begin{equation}
 ds^2=\left(1+ \frac{2\phi}{c^2}\right) c^2 \, dt^2 - d\mathbf{x}^2
 \label{dssquare}
\end{equation} 
where $\phi$ is the Newtonian potential. 
The Feynman propagator again has the integral representation
\begin{equation}
 G(x,y) = \int_0^\infty ds\, e^{-im^2 s}\, K(x,s; y,0) =  \int_0^\infty \frac{ds}{s^2} \sum_{paths} e^{-im^2 c^2s/\hbar^2}\, e^{-il^2(x,y)/4s}
\end{equation}
where $l^2$ now represents the proper length of a path between two events evaluated using the metric in \eq{dssquare}. Assuming that we can formally interchange the sum over paths and the integration over $s$, we can express the above result as 

\begin{eqnarray}
 G(x,y) &=& \sum_{\rm paths} \int_0^\infty ds\ e^{-im^2c^2s/\hbar^2}  \exp\left\{ -\frac{i}{4} \int_0^s d\tau \left[c^2 \left( 1+\frac{2\phi}{c^2}\right) \dot t^2 - 
 \dot{\mathbf{x}}^2\right]\right\}\nonumber\\
&=&  \sum_{\rm paths} \int_0^\infty ds\ e^{-im^2c^2s/\hbar^2} \exp\left\{ -\frac{ic^2}{4} \int_0^s d\tau \ \dot t^2 \right\} F(s)
\end{eqnarray}
where 
\begin{equation}
F(s) = \exp\left\{ \frac{i}{4} \int_0^s d\tau \ \dot{\mathbf{x}}^2 \right\}\exp\left\{ -\frac{i}{2} \int_0^s d\tau \phi \ \dot{t}^2 \right\}                                                                                                                                                   \end{equation} 
is treated as a function of the parameter $s$.
Thus, we can write the expression for the propagator as:
\begin{equation}
G(x,y) =  \int\mathcal{D}\mathbf{x}\ \mathcal{D}t   \int_0^\infty ds\ e^{-im^2c^2s/\hbar^2} \exp\left\{ -\frac{ic^2}{4} \int_0^s d\tau \ \dot t^2 \right\} F(s)  
\label{propg}
\end{equation} 
where we have identified the sum over paths as functional integrals over spatial and time coordinates:
\begin{equation}
\sum_{\rm paths} \equiv \int\mathcal{D}\mathbf{x}\ \mathcal{D}t
\end{equation} 

In order to estimate the above integral in the $c \to \infty$ limit, we note that this limit is equivalent to taking a saddle point on the integrand treated as a function of the parameter $s$. The saddle point of the integrand is given by:

\begin{equation}
 \frac{m^2c^2}{\hbar^2} = \frac{c^2}{4}\left. \left( \frac{dt}{d\tau}\right)^2 \right|_{\tau=s} \Rightarrow t = \frac{2ms}{\hbar}
 \label{saddleint}
\end{equation} 
We now evaluate the  integral in \eq{propg} in the saddle point limit given by \eq{saddleint}. The functional integral over all paths  $\mathcal{D}t$ reduces, in saddle point, to the condition that  $\dot t$ = constant. From our preceding argument for the saddle point of $s$, we see that the value of this constant is $(2 m/\hbar)$. 
If we use these results in the expression for $G(x,y)$, we find:

\begin{eqnarray}
 G(x,y) &=&\int \mathcal{D} \mathbf{x} \, \exp \left[ - \frac{im c^2 t}{2 \hbar}\right] \, \exp\left[ -\frac{ic^2}{4 } \frac{\hbar t}{2 m }\left(\frac{4m^2}{\hbar^2}\right)\right] \nonumber\\
 &&\qquad\qquad \qquad \times \qquad \exp\left[ - \frac{i}{4}\int_0^s d\tau \left( 2\phi \ \dot t^2 - \dot{\mathbf{ x}}^2\right)\right]\nonumber\\
&=& \int \mathcal{D}\mathbf{x }\, e^{-imc^2t/\hbar} \exp\left\{ -\frac{i}{4} \int_0^s d\tau\, \dot t^2 \left[ 2\phi - \left(\frac{d\mathbf{x}}{dt}\right)^2\right]\right\} 
\end{eqnarray} 
With these, admittedly formal, manipulations we have  succeeded in converting a four-dimensional functional  integral (over $t$ and $\mathbf{x}$) and an ordinary integral over $s$  into the standard three-dimensional functional integral over $\mathbf{x}$ which is appropriate for the non-relativistic limit.

According to our previous arguments, the validity condition for the saddle-point approximation is that $\dot t$ is a constant, and hence, it can be taken out of the integral in the phase of the exponential, and substituted by $(2m/\hbar)$. We also change the variable of integration form $\tau$ to $t$ (noting that the entire operation is performed under the condition that, at $\tau = s$,  $(dt/d\tau) = 2 m/\hbar$). We get:

\begin{eqnarray}
 G(x,y)&=&\int \mathcal{D}\mathbf{x }\, e^{-imc^2t/\hbar}\exp \left\{ -\frac{i}{4} \left( \frac{4m^2}{\hbar^2}\right) \int_0^{t} \frac{\hbar dt}{2m}\left[ 2\phi - \left(\frac{d\mathbf{x}}{dt}\right)^2\right]\right\} \nonumber\\
 &=& \int \mathcal{D}\mathbf{x }\, e^{-imc^2t/\hbar}\exp \left\{ -\frac{im}{\hbar}  \int_0^{t} \phi \, dt+ \int_0^{t} \frac{im}{2\hbar} \left(\frac{d\mathbf{x}}{dt}\right)^2 dt \right\} \nonumber\\
 &=& \int \mathcal{D}\mathbf{x }\, e^{-imc^2t/\hbar}\exp \left\{ \frac{i}{\hbar}\left[  \int_0^{t}\frac{m}{2} \left(\frac{d\mathbf{x}}{dt}\right)^2 dt -  \int_0^{t} m\,\phi\, dt  \right]\right\}
\end{eqnarray}

The above expression is identical in form (except, of course, for the standard $\exp{-imc^2t/\hbar}$ phase) to the non-relativistic Feynman path integral for a particle of mass $m$ in an external potential $\phi$.  This proves that in the $c \to \infty$ limit, the Feynman propagator does indeed reduce to the non-relativistic Feynman path integral for the case of an external weak gravitational potential.

\section{Lack of transitivity of the Feynman propagator in relativistic field theory}

In non-relativistic quantum mechanics, the kernel satisfies the transitivity condition given by \eq{ten}. This represents the fact that a particle which propagates from $(x_1, t_1)$ to $(x_3, t_3)$ must exist at some unique spatial location $x_2$ at any intermediate time $t_2$. We cannot have  a similar transitivity condition obeyed in terms of spatial coordinates, which is apparent from the fact that time and space coordinates enter differently in the kernel. In the path integral, one evaluates paths which go forward in time but can go forward or backward in spatial coordinates.

In the case of the Feynman propagator, we instead work with the path integral in \eq{twenty} in which the particles go forward in the \textit{proper time} $s$, but can go forward or backward both in coordinate time and spatial coordinates. After evaluating the path integral, we do a Fourier transform with respect to the proper time to obtain the Feynman Green's function. It is obvious that, just as we cannot have transitivity vis-a-vis spatial coordinates in the non-relativistic kernel, we cannot have transitivity in spatial \textit{or} time coordinates for the Feynman propagator. One can, in fact, prove that the transitivity integral for the Feynman propagator leads to the result 
\begin{eqnarray}
 \int_{-\infty}^\infty G_F(x_3; x_2) G_F(x_2; x_1) d^4 x_2 &=& i\frac{\partial}{\partial(m^2)} G_F(x_3; x_1)\nonumber\\
 &=& G_F(x_3; x_1)\left[ i\frac{\partial}{\partial (m^2)} \ln  G_F(x_3; x_1)\right]
 \label{twentynine} 
\end{eqnarray} 
where the  factor in square brackets expresses the lack of transitivity.
We shall provide a simple demonstration of this result since we have not seen it discussed in the literature. 

The result can be obtained from the Schwinger representation of the propagator by straightforward integration (as shown in Appendix \ref{app6}). However, it is somewhat easier to prove the result using the well known momentum space representation of the Feynman propagator given by
\begin{equation}
 G_F(x,y) = i \int \frac{d^4p}{(2\pi)^4 } \frac{e^{-ip(x-y)}}{(p^2-m^2+i\epsilon)}
 \label{thirty}
\end{equation} 
Explicit substitution of \eq{thirty} in \eq{twentynine} gives:
\begin{eqnarray}
 &&\quad \int_{-\infty}^\infty G_F(t_3,\mathbf{x}_3; t_2,\mathbf{x}_2) G_F(t_2,\mathbf{x}_2; t_1,\mathbf{x}_1) d^4 x_2\nonumber\\
 && \qquad\qquad=\int_{-\infty}^\infty d^4 x_2 \left[ \frac{d^4p}{(2\pi)^4} \, \frac{i e^{-ip(x_3-x_2)}}{p^2 - m^2+ i\epsilon}\right]\left[ \frac{d^4k}{(2\pi)^4} \, \frac{i e^{-ik(x_2-x_1)}}{k^2 - m^2+ i\epsilon}\right]\nonumber\\
 && \qquad\qquad= -\frac{1}{(2\pi)^4}\int  \frac{d^4p \,  e^{-ip(x_3-x_1)}}{(p^2 - m^2+ i\epsilon)^2}
\end{eqnarray} 
Since
\begin{equation}
 \frac{\partial}{\partial(m^2)} G_F(x_3, x_1) =i \int \frac{d^4p}{(2\pi)^4 } \frac{e^{-ip(x_3-x_1)}}{(p^2-m^2+i\epsilon)^2}
\end{equation} 
we immediately get the result in \eq{twentynine}.
Once again, in the non-relativistic limit, the extra factor in \eq{twentynine} reduces to unity which can be demonstrated by straightforward manipulation. It is also interesting to notice that the kernel in \eq{kcorrection}, which contains the first-order relativistic correction to the non-relativistic path integral, does not possess the transitivity property due to the presence of additional terms in the integrand. 

The origin of this non-transitivity can be understood from the fact that, \textit{even in the non-relativistic case}, the \textit{energy kernel} is not transitive. This is to be expected purely from dimensional analysis and can be seen as follows.
The energy kernel in non-relativistic quantum mechanics is given by: 
\begin{equation}
 G(E, \mathbf{x}_2,\mathbf{x}_1) = \int_{-\infty}^\infty dT\ e^{iET}\, K(T,\mathbf{x}_2,\mathbf{x}_1)
\end{equation} 
so that $G$ has the dimensions of $[TL^{-3}]$. If $G$ is to be transitive, we must have
\begin{equation}
 \int d^3 \mathbf{x}_2 \, G(E, \mathbf{x}_3,\mathbf{x}_2) G(E,\mathbf{x}_2,\mathbf{x}_1) \stackrel{?}{=} G(E, \mathbf{x}_3, \mathbf{x}_1)
 \label{gtrans}
\end{equation} 
which is clearly impossible since the left-hand side has dimensions of $[L^{-3}T^2]$ and the right-hand side, that of $[TL^{-3}]$. Clearly, one needs a factor having dimensions of time (or equivalently, of inverse energy) on the right-hand side of \eq{gtrans} to make it dimensionally consistent.

The correct result  can be easily obtained  when the action has no explicit time dependence. The non-relativistic energy kernel can be expressed as:
\begin{eqnarray}
 G(E, \mathbf{x}_2, \mathbf{x}_1) &=& \int_{-\infty}^{\infty} dT e^{iET} K(T, \mathbf{x}_2, \mathbf{x}_1) = \int_{0}^{\infty} dT e^{iET} \sum_{n}e^{-i E_n T} \phi_n(\mathbf{x}_2) \phi_n^*(\mathbf{x}_1) \nonumber \\
 &=& \sum_{n} \frac{i \phi_n(\mathbf{x}_2) \phi_n^*(\mathbf{x}_1)}{E - E_n + i\epsilon} 
\end{eqnarray} 
where $\phi_n$ is the energy eigenfunction corresponding to the eigenvalue $E_n$.
Hence, we get, on using the orthonormality of energy eigenfunctions:
\begin{eqnarray}
\label{gentrans}
 \int d^3\mathbf{x} \ G(E, \mathbf{x}_2, \mathbf{x}) G(E, \mathbf{x}, \mathbf{x}_1) &=& -\sum_{n} \frac{1}{(E - E_n + i\epsilon)^2}\phi_n(\mathbf{x}_2)\phi_n(\mathbf{x}_1) \nonumber\\
                                   &=& -i \frac{\partial G}{\partial E} =  G\left[-i \frac{\partial}{\partial E} \ln G\right]
\end{eqnarray} 
which is similar in form to the result we obtained above in \eq{twentynine} for the Feynman propagator.  In fact, the equivalence arises  from the fact that the Feynman propagator is just the energy Fourier transform (with $E$ replaced by $mc^2$) of the path-integral with the relativistic quadratic action, as is clear from \eq{twenty}. (The sign flip between the right-hand sides of \eq{gentrans} and \eq{twentynine} is unimportant, being the result of the difference in the Fourier transform conventions used in their derivations.) Hence, the non-transitivity of the Feynman propagator can be understood directly from this fact, and the result does not require  any deeper quantum field theoretic concepts (like that of antiparticles) for its explanation. 

\section{Conclusions}

We have discussed several aspects of  the non-relativistic limit of quantum field theory in both inertial and noninertial frames and in weak gravitational fields. Starting from curious subtleties in the transition from special relativity to classical mechanics, we have explained the physical origin of the extra term in the non-relativistic action under Galilean transformations, thereby clarifying some points discussed previously in the literature. We have taken the non-relativistic limit of the generally covariant Klein-Gordon equation, and shown that it reduces to the Schrodinger equation in the presence of the corresponding gravitational field, thus providing an interpretation of the principle of equivalence in quantum mechanics. We also showed how such limiting processes can be described, fairly rigorously, in the language of propagators, which is essential because relativistic field theory does not allow an interpretation in terms of a single particle wave function. 
Along the way, we have indicated how  the relativistic Feynman propagator can be explicitly obtained from the standard special relativistic square-root action functional and connected it up with the Schwinger proper time description. We have also obtained the non-relativistic limit of the Feynman propagator in the inertial and Rindler frames, and in a weak gravitational field, and demonstrated its reduction to the corresponding path integral kernels. Thus, we have effectively explored the GQM vertex of the physics cube (Fig.\ref{fig:physicscube}), and studied the $c^{-1}=0$ plane as a projection from other parts of the cube. While some of these results are obvious with hindsight, we were persuaded to present the details because we have not seen them discussed properly in textbooks or in published literature.

As Fig.\ref{fig:physicscube} shows, the exact theories of physics allow useful, but approximate, limiting forms. Sometimes, these approximate versions --- though incorrect --- can be internally consistent with all their features  being understandable within the approximation. But in some other contexts, approximate theories can host features which are residues of the more exact description. This paper highlights some features of the nonrelativistic limit which belong to the latter class. We saw that the behaviour of the action functional and the phase of the wave function in the nonrelativistic limit have no simple explanation within the nonrelativistic theory. But they emerge very naturally from the nonrelativistic limit of the term $mc^2(\tau -t)$,
thereby acquiring a direct physical meaning in the relativistic domain. This interpretation, as we saw, holds even in a noninertial frame and in the presence of a weak gravitational field. What is curious is the fact that this phase has a nonzero, finite, limit even when $c\to\infty$.
This is essential to make quantum mechanics and Galilean relativity consistent; essentially, one needs to start with relativistic quantum theory and Lorentz transformations, and take appropriate limits to understand what happens when one combines quantum mechanics with Galilean relativity!
This is an example of a more exact theory leading to curious residues in its approximate version.

\section*{Acknowledgements} This paper was written when one of the authors (T.P.) was visiting the Institute of Astronomy, Cambridge and the hospitality of the IOA is gratefully acknowledged. The research of T.P is partially supported by the J.C. Bose Fellowship of DST, India. The research of H.P is  supported by the KVPY Fellowship of DST, India. We thank D.V. Ahluwalia and C.S. Unnikrishnan for detailed comments on the manuscript.

\appendix

\section{}
\label{app2}
In this appendix we prove that the wave function \eq{wavefntrans} satisfies the Schrodinger equation, \eq{schrotrans}. We set $m = \hbar = 1$ for convenience, so that \eq{schrotrans} becomes:
\begin{equation}\label{schrotransmh1}
 i\frac{\partial\Psi}{\partial t} = - \frac{1}{2} \frac{\partial^2\Psi}{\partial x^2} - \ddot \xi x \Psi
\end{equation} 
The co-ordinate transformation is given by $x' = x - \xi(t),\  t' = t$. 
We now substitute $\Psi(x,t) = \Psi' (x',t') e^{-if}$ into the above equation, where $f =  -x \dot \xi + \frac{1}{2} \int \dot \xi^2 dt$. We have the following relations:

\begin{equation}
 i\frac{\partial (\Psi' e^{- if})}{\partial t} = i\frac{\partial \Psi'}{\partial t} e^{- if} +  e^{- if}\Psi'\frac{\partial f}{\partial t}
\end{equation} and 
\begin{equation}
 \frac{\partial}{\partial x}(\Psi' e^{- if}) =  e^{- if}\frac{\partial \Psi'}{\partial x'} - i e^{- if}\frac{\partial f}{\partial x}\Psi'
\end{equation} 
Hence,
\begin{equation}
 \frac{\partial^2}{\partial x^2}(\Psi' e^{- if}) = e^{-if} \frac{\partial^2 \Psi'}{\partial x'^2} - 2i\frac{\partial \Psi'}{\partial x'}\frac{\partial f}{\partial x} e^{-if} - e^{- if}\left(\frac{\partial  f}{\partial x}\right)^2 \Psi' 
\end{equation} 
where we have used the facts that $\partial \Psi'/\partial x = \partial \Psi'/\partial x'$ and $\partial^2 f/\partial x^2 = 0.$
Using these relations, \eq{schrotransmh1} becomes:
\begin{eqnarray}\label{schrotransmh12}
 i\frac{\partial \Psi'}{\partial t}  +  \Psi' \frac{\partial f}{\partial t} &=& -\frac{1}{2}  \frac{\partial^2 \Psi'}{\partial x'^2}  + i\frac{\partial \Psi'}{\partial x'}\frac{\partial f}{\partial x} \nonumber \\&+& \frac{1}{2}\left(\frac{\partial  f}{\partial x}\right)^2 \Psi' - \ddot{\xi} \Psi' x
\end{eqnarray}  
We also know that
\begin{equation}
 \frac{\partial \Psi'}{\partial t}=\frac{\partial \Psi'}{\partial t'} - \dot \xi \frac{\partial \Psi'}{\partial x'}
\end{equation} 
and
\begin{equation} 
 \frac{\partial f}{\partial x} =  -\dot \xi; \quad  \frac{\partial f}{\partial t }=  -\ddot{\xi} x  + \frac{1}{2}\dot{\xi}^2
\end{equation}
Using the above relations in \eq{schrotransmh12}, it readily transforms to:
\begin{equation}
 i\frac{\partial \Psi'}{\partial t'} =  - \frac{1}{2}\frac{\partial^2 \Psi'}{\partial x'^2}
\end{equation} 
which is satisfied identically, since we know that $\Psi'(x',t')$ is a solution to the free particle Schrodinger equation in the $(x',t')$ frame of reference. Hence, we see that the wave function in \eq{wavefntrans} satisfies \eq{schrotrans}.

\section{}
\label{app3}

In this appendix we determine the scaling of the energy levels $E_N$ of the (in general) non-Hookean oscillator with the quantum number $N$. Consider \eq{bswkb}  which, for the case of the power-law potential (\ref{powerlawpot}), becomes:
\begin{equation}
 \int (2 m (E_N - \alpha|x|^n))^{1/2} dx = N \hbar
\end{equation} 
This integral is to be evaluated between the two turning points. As the potential is symmetric in $x$, we can write:
\begin{equation}
 N \propto \sqrt{m E_N} \int_{0}^{(E_N/a)^{1/n}} \left(1 - \frac{\alpha x^n}{E_N} \right)^{1/2} dx
\end{equation} 
Substituting $y = \alpha x^n/E_N$, we get:
\begin{equation}
 N \propto m^ {1/2} E_N^ {(n + 2)/2 n} \left(\frac{\alpha^{-1/n}}{n}\right) \int_0^1 \frac{dy}{y^{(n-1)/n}} \sqrt{1 - y}
\end{equation} 
Since we are only interested in the scaling of $E_N$ with $N$ and $m$, we can see that (using the fact that $\alpha \propto m$):
\begin{equation}
 E_N \propto N^{2n/(n+2)} \ m^{(n-2)/(n+2)}
\end{equation}
Hence, we see that, only in the case $n = 2$, we have $E_N \propto N$, with the mass dependence scaling out. In all other cases, the energy levels $E_N$, and consequently, the transition frequency between successive energy levels, will depend on $m$, the mass of the oscillator.
[The full expression for $E_N$ in the WKB approximation has been worked out in, for e.g., \cite{sukhatme}.]

\section{}
\label{app4}

The result in \eq{nineteen} can be proved by using the Hamiltonian form of the Feynman path integral and converting the path integral sums over the coordinates and momenta in an appropriate form.
We will need to use the following three standard results in functional integration: 
\begin{equation}
  \delta (f(t)) = \sum_{\lambda(t)} \exp i \int dt\ \lambda(t) f(t)
  \label{aone}
 \end{equation} 

\begin{equation}
 \sum_{\mathbf{p}} \exp i \int dt\, \left[ \mathbf{p \cdot \dot x} + a(t) p^2\right] = \exp i \int dt\, \frac{\dot x^2}{4 a (t)}
 \label{atwo}
\end{equation}

\begin{equation}
  \sum_{\lambda(t)} \exp i \int dt\ \left( \lambda(t) a(t) +\frac{b(t)}{\lambda(t)}\right) = \exp i \int dt \left[ - 4 ab\right]^{1/2}
  \label{athree}
\end{equation} 
The \eq{aone} is just the definition of delta functional; \eq{atwo} and \eq{athree} can be easily obtained in the Euclidean sector by usual time-slicing techniques and can be analytically continued. They represent direct generalizations of the corresponding results  for ordinary integrals.

To prove \eq{nineteen}, we begin by writing the standard Feynman path integral in the Hamiltonian form:
 \begin{equation}
 K({\bf x_2, x_1};T)=
\sum_{all{\bf x}}
\sum_{all{\bf p}}\exp
{i\over \hbar}\int^T_0
dt
({\bf p.\dot x}-H({\bf p,x})); \quad H = \frac{\mathbf{p}^2}{2m_0} + V(\mathbf{x})
\label{thirteen}
\end{equation} 
 where the sum is over all functions $\mathbf{p}(t)$ but only those functions $\mathbf{x}(t)$ which satisfy the boundary conditions. (We will assume $H\geq0$.) We now introduce inside the functional integral, the factor unity in the form:
 \begin{equation}
 1 = \int_0^\infty d E \, \delta (E- H(\mathbf{p,x}))
\end{equation} 
leading to:
 
\begin{eqnarray}
 K(\mathbf{x},t) &=& \int_0^\infty d E \,\sum_{\mathbf{x}} \sum_{\mathbf{p}}  \delta (E- H(\mathbf{p,x}))
  \exp \frac{i}{\hbar} \int dt\, \left( \mathbf{p \cdot 
\dot x} - H (\mathbf{p,x}\right)\nonumber\\
&=&\int_0^\infty d E \,\sum_{\mathbf{x}} \sum_{\mathbf{p}}  \delta (E- H) e^{-iEt} \, \exp i \int dt\, (\mathbf{p\cdot \dot x}) 
\end{eqnarray} 
So
\begin{equation}
\int_0^\infty K(\mathbf{x},t)    e^{iEt} dt \equiv B(\mathbf{x}, E) =   \sum_{\mathbf{x}} \sum_{\mathbf{p}} \delta \left( \frac{\mathbf{p}^2}{2m} + V(\mathbf{x}) - E\right)  \exp i \int dt\, (\mathbf{p\cdot \dot x})  
\label{defB}
\end{equation} 
We now express the delta functional using \eq{aone}:
\begin{equation}
 \delta \left( \frac{\mathbf{p}^2}{2m} + V(\mathbf{x}) - E\right) =  \sum_{\lambda(t)} \exp i \int  \lambda(t) \left[\frac{\mathbf{p}^2}{2m} + V(\mathbf{x}) - E\right] dt
\end{equation}  
We then get, by straightforward manipulations, the result: 
 
\begin{eqnarray}
 B(\mathbf{x},E) &=& \sum_{\mathbf{x}}\sum_{\lambda(t)} \exp i \int dt\,\lambda(t)  \left[ V(\mathbf{x}) - E\right] \sum_{\mathbf{p}} \exp i\int dt\,\left[ \mathbf{p \cdot \dot x} + \frac{\lambda(t)}{2m} p^2\right]\nonumber\\
 &=& \sum_{\mathbf{x}}\sum_{\lambda(t)} \exp i \int dt\, \lambda(t)\left[ V(\mathbf{x}) - E\right] \exp i\int dt\,\frac{1}{2} \frac{m}{\lambda(t)}\mathbf{\dot{x}}^2\nonumber\\
&=& \sum_{\mathbf{x}}\sum_{\lambda(t)} \exp i \int dt\, \left[\lambda(t) \left( V(\mathbf{x}) - E\right) + \frac{m}{2\lambda(t)} \mathbf{\dot{x}}^2\right]\nonumber\\
&=& \sum_{\mathbf{x}} \exp i \int dt\, \sqrt{ 2m (E-V)\, |\mathbf{\dot{x}}|^2}
\end{eqnarray}
The left hand side is just the propagator for a particle with energy $E$, which now has an explicit path integral representation in terms of a square-root action. 
Using this result in \eq{defB}, we obtain \eq{nineteen}.

\section{}
\label{app5}

In this appendix, we describe how the non-relativistic \textit{kernel} transforms when one goes over from an inertial frame to an  accelerating frame, and use our general result to verify explicitly the form of the kernel for the case of a particle moving with uniform acceleration.

 We consider the two frames of reference $S$ and $S'$, where the particle is free in $S' = (x',t')$ and accelerated in the frame $S = (x, t)$; $S$ and $S'$ being related by the transformation $x' =  x- \xi(t)$; with $ t' = t$.  Let $K'(x_b', t_b; x_a', t_a)$ be the non-relativistic propagator in $S'$, where $(x_b', t_b)$ and $(x_a', t_a)$ are the co-ordinates of the final and initial points respectively. Then, it turns out that in the frame $S$, in which the particle has an acceleration $\ddot \xi$, the propagator $K(x_b, t_b; x_a, t_a)$ takes the form:
\begin{equation}
\label{ktrans}
 K(x_b, t_b; x_a, t_a) = K'(x_b', t_b; x_a', t_a) e^{-i \Lambda/\hbar}
\end{equation} 
where 
\begin{equation}
\Lambda \equiv (-m x_b \dot{\xi_b} + m x_a \dot{\xi_a}) +(1/2) m \int_{t_a}^{t_b} \dot{\xi}^2 dt
\end{equation}
Here, the overdot denotes differentiation with respect to time, and an expression like $\dot{\xi_b}$ denotes the time derivative of $\xi$ evaluated at $t_b$.

The above relationship is easy to understand along the following lines:
Recall that  the Lagrangian picks up a total time derivative term as we go from frame $S'$ to frame $S$ (or vice-versa). Since we are dealing with a free particle, the Lagrangian is a quadratic function of the co-ordinates and velocities, and the transformation preserves the quadratic nature of the Lagrangian. We know that, for quadratic Lagrangians, the propagator contains a phase term that is proportional to the action functional for the classical trajectory between the initial and the final points. From \eq{lagtrans}, we see that $\mathcal{A}$ and $\mathcal{A'}$, the classical action functionals in the two frames,  are related by $\mathcal{A'} = \mathcal{A} + f(x,t)$. Hence,  we would expect that the propagators in the same two frames are related by  $K(x,t) = K'(x',t) \exp[-i (f_b - f_a)/\hbar]$ where $f_b \equiv f(x_b, t_b)$ and $f_a \equiv f(x_a, t_a)$. From \eq{fform} we can see that  
\begin{equation}
(f_b - f_a) = m(x_a\dot{\xi_a} - x_b\dot{\xi_b}) +  (1/2) \int_{t_a}^{t_b} m \dot{\xi}^2 dt \equiv \Lambda,                                                                                                           \end{equation}  
from which the result (\ref{ktrans}) follows. 
 
We now specialize to the case of uniform acceleration. We have a free particle in the frame of reference $S' = (x', t')$ and we transform from this frame to frame $S = (x,t)$ where 
$x' = x - (1/2) \ddot{\xi}t^2$ , $t' = t$ , and $\ddot{\xi}$ is a constant. Consider first the free particle kernel $K'$ in the frame $(x',t')$ which is given by:
\begin{equation}
\label{propfree}
 K'(x'_b, t_b; x'_a, t_a) = N(t_b, t_a) \exp \left(\frac{i m}{\hbar}\frac{\left(x_b' - x_a' \right)^2}{(t_b - t_a)}\right)
\end{equation} where $N$ is a normalization constant, which depends on the initial and final time co-ordinates $t_a$ and $t_b$ respectively, and we have used the fact that $t' = t$.
Since we have translational invariance as far as the time co-ordinate is concerned, we can replace $t_a$ and $t_b$ by $0$ and $T$ respectively.

We now substitute for the co-ordinate $x'$ in terms of $x$ and also add the phase factor $e^{-i\Lambda/\hbar}$. Here, $\Lambda$ is given by:
\begin{eqnarray}
\label{lambdaunig}
 \Lambda &\equiv&  m(x_a\dot{\xi_a} - x_b\dot{\xi_b}) +  \frac{1}{2}\int_{0}^{T} m \dot{\xi}^2 dt
 \nonumber \\ &=& m\left(-x_b\ddot{\xi}T + \frac{1}{6}\ddot{\xi}^2 T^3\right)
\end{eqnarray} 
Substituting \eq{propfree} and \eq{lambdaunig} into \eq{ktrans}, we find:
\begin{eqnarray}
 K(x_b, T; x_a, 0) &=& N(T)
  \exp \frac{i m}{2 \hbar T }  \left[\left(x_b - \frac{1}{2}\ddot{\xi}T^2\right)^2\right. \nonumber\\
  &&  \left. + \ x_a^2 - 2 x_a \left(x_b - \frac{1}{2} \ddot{\xi} T^2\right) + 2 x_b\ddot{\xi}T^2 - \frac{1}{3}\ddot{\xi}^2 T^4 \right] 
\end{eqnarray} 
On simplification, we get:
\begin{equation}
\label{propuniaccn}
 K(x_b, T; x_a, 0) = N(T) \exp \frac{i m }{2 \hbar T }  \left((x_b - x_a)^2 - \frac{1}{12}\ddot{\xi}^2 T^4 + \ddot{\xi} T^2( x_b + x_a)\right)
\end{equation} 
which matches identically with the standard expression for the propagator for a uniformly accelerated particle moving with acceleration $\ddot{\xi}$ \cite{feynmanhibbs}.  Hence, we have verified the validity of our general formula \eq{ktrans} for the special case of the frame $(x,t)$ being a uniformly accelerated frame of reference.

\section{}
\label{app6}

We provide a direct proof of \eq{twentynine} working in the Euclidean sector and evaluating the integral on the left hand side explicitly.
From the expression for the Euclidean Feynman propagator: 
\begin{equation}
 G_F(z;y) = \frac{1}{(4\pi)^2} \int_0^\infty \frac{d\lambda}{\lambda^2} e^{-\lambda m^2} e^{-(z_E-y_E)^2/4\lambda}
\end{equation} 
in the left hand side of \eq{twentynine}, we get: 
\begin{eqnarray}
 \int_{-\infty}^\infty G_F(z;y)\, G_F(y;x)d^4y
 &=& \frac{1}{(4\pi)^4} \int_{-\infty}^\infty d^4y\,e^{-(z_E-y_E)^2/4\lambda} e^{-(y_E-x_E)^2/4\mu}\nonumber\\
 &&\qquad \qquad \times
 \int_0^\infty\int_0^\infty \frac{d\lambda}{\lambda^2}\frac{d\mu}{\mu^2} e^{-(\lambda+\mu)m^2}
\end{eqnarray} 
Now, using the standard result:
\begin{equation}
\int_{-\infty}^\infty \sqrt{\frac{a}{\pi}}\sqrt{\frac{b}{\pi}} \, e^{-a(x-\mu)^2} \, e^{-b(u-y)^2} du 
=\sqrt{\frac{ab}{\pi(a+b)}} \, \exp\left(- \frac{ab}{a+b}\, (x-y)^2\right)
\end{equation} 
we find that 
\begin{equation}
 \int_{-\infty}^\infty G_F(z;y)\, G_F(y;x)d^4y = -\frac{i}{(4\pi)^2}
\int_{0}^\infty \frac{d\lambda \ d\mu \, e^{-(\lambda+\mu) m^2}}{(\mu+\lambda)^2}\, \exp\left[ -\frac{1}{4(\lambda+\mu)}\, (x_E-z_E)^2\right]
\end{equation} 
Transforming the variables to $\alpha = \mu + \lambda$ and $\beta=\mu -\lambda$ gives:
\begin{eqnarray}
 \int_{-\infty}^\infty G_F(z;y) G_F(y;x)\, d^4y &=& -\frac{i}{2(4\pi)^2}\int_{0}^\infty \int_{-\alpha}^\alpha\frac{d\beta\, d\alpha}{\alpha^2} e^{-\alpha m^2} \, e^{-(x_E-z_E)^2/4\alpha}\nonumber\\
 &=& -\frac{i}{(4\pi)^2} \int_{0}^\infty \frac{d\alpha}{\alpha}\, e^{-\alpha m^2}\,  e^{-(x_E-z_E)^2/4\alpha}\nonumber\\
 &=& i\frac{\partial}{\partial(m^2)} G_F(z; x)
\end{eqnarray} 
This demonstrates the result in \eq{twentynine}.

\end{document}